\documentclass[twocolumn,twoside]{IEEEtran}
\usepackage{mathpazo}
\usepackage{times}
\usepackage{amsmath}
\usepackage{amsfonts}
\usepackage{latexsym}
\usepackage{amssymb}

\usepackage{upref}
\usepackage{theorem}
\usepackage{graphicx}
\usepackage{psfrag}
\usepackage[noadjust]{cite}
\usepackage{color}
\usepackage{overpic}
\usepackage{rotating}
\usepackage{mathtools}
\usepackage{xparse}
\DeclarePairedDelimiterX{\set}[1]{\{}{\}}{\setargs{#1}}
\DeclarePairedDelimiterX{\cond}[1]{[}{]}{\setargs{#1}}
\NewDocumentCommand{\setargs}{>{\SplitArgument{1}{;}}m}
{\setargsaux#1}
\NewDocumentCommand{\setargsaux}{mm}
{\IfNoValueTF{#2}{#1} {#1\,\delimsize|\,\mathopen{}#2}}

\DeclarePairedDelimiter\abs{\lvert}{\rvert}
\DeclarePairedDelimiter\ceil{\lceil}{\rceil}
\DeclarePairedDelimiter\floor{\lfloor}{\rfloor}
\DeclarePairedDelimiter\parenv{\lparen}{\rparen}
\DeclarePairedDelimiter\sparenv{\lbrack}{\rbrack}

\DeclarePairedDelimiter\spn{\langle}{\rangle}

\theoremstyle{plain}
\newtheorem{theorem}{Theorem$\!$}
\newtheorem{lemma}[theorem]{Lemma$\!$}
\newtheorem{proposition}[theorem]{Proposition$\!$}
\newtheorem{corollary}[theorem]{Corollary$\!$}
\newtheorem{definition}[theorem]{Definition$\!$}
\newtheorem{example}[theorem]{Example$\!$}
\newtheorem{remark}[theorem]{Remark$\!$}

\newcommand{\cL}{\mathcal{L}}

\renewcommand{\leq}{\leqslant}

\renewcommand{\geq}{\geqslant}

\renewcommand{\Bbb}{\mathbb}

\newcommand{\Cref}[1]{Co\-ro\-lla\-ry\,\ref{#1}}

\renewcommand{\Bbb}{\mathbb}

\newcommand{\F}{\mathbb{F}}

\newcommand{\N}{{\Bbb N}}

\newcommand{\R}{{\Bbb R}}

\newcommand{\Prob}{{\Bbb P}}

\newenvironment{bsmatrix}{\left[\begin{smallmatrix}}{\end{smallmatrix}\right]}

\newcommand{\eqdef}{\triangleq}
\newcommand{\sbinom}[2]{\genfrac{[}{]}{0pt}{}{#1}{#2}}

\newcommand{\oC}{\overline{C}}
\newcommand{\oc}{\overline{c}}
\newcommand{\oD}{\overline{D}}
\newcommand{\bc}{\mathbf{c}}
\newcommand{\tc}{\widetilde{c}}
\newcommand{\ove}{\overline{e}}
\newcommand{\og}{\overline{g}}
\newcommand{\oG}{\overline{G}}
\newcommand{\oh}{\overline{h}}
\newcommand{\bv}{\mathbf{v}}
\newcommand{\ov}{\overline{v}}
\newcommand{\bu}{\mathbf{u}}
\newcommand{\ou}{\overline{u}}

\newcommand{\ow}{\overline{w}}

\newcommand{\os}{\overline{s}}
\newcommand{\ox}{\overline{x}}
\newcommand{\oy}{\overline{y}}
\newcommand{\bz}{\mathbf{z}}
\newcommand{\oz}{\overline{z}}
\newcommand{\ozero}{\overline{0}}
\newcommand{\oone}{\overline{1}}
\DeclareMathOperator{\supp}{supp}
\DeclareMathOperator{\gl}{GL}
\DeclareMathOperator{\wt}{wt}

\DeclareMathOperator{\rank}{rank}
\DeclareMathOperator{\Var}{Var}
\DeclareMathOperator{\Cov}{Cov}
\DeclareMathOperator{\E}{E}
\definecolor{dgreen}{rgb}{0.0, 0.5, 0.0}

\newcommand{\ind}{\mathbb{I}}
\newcommand{\zero}{\mathbf{0}}

\outer\def\proclaim #1. #2\par{\medbreak
 \noindent{\bf#1.\enspace}{\sl#2\par}%
 \ifdim\lastskip<\medskipamount \removelastskip\penalty55\medskip\fi}

\begin{document}

\title{\textbf{The Generalized Covering Radii of \\ Linear Codes}}

\author{
  Dor Elimelech~\IEEEmembership{Student Member,~IEEE}, Marcelo Firer~\IEEEmembership{Member,~IEEE}, and Moshe Schwartz~\IEEEmembership{Senior Member,~IEEE}%
  \thanks{This work was submitted in part to the IEEE International Symposium on Information Theory (ISIT) 2021.}%
  \thanks{Dor Elimelech is with the School
    of Electrical and Computer Engineering, Ben-Gurion University of the Negev,
    Beer Sheva 8410501, Israel
    (e-mail: doreli@post.bgu.ac.il).}%
  \thanks{Marcelo Firer is with the Institute of Mathematics, Statistics and Scientific Computing, University of Campinas, Campinas 13083-859, Brazil (e-mail: mfirer@ime.unicamp.br).}%
  \thanks{Moshe Schwartz is with the School
    of Electrical and Computer Engineering, Ben-Gurion University of the Negev,
    Beer Sheva 8410501, Israel
    (e-mail: schwartz@ee.bgu.ac.il).}%
  \thanks{The work of D. Elimelech was supported in part by an Israel Science Foundation (ISF) Grant under Grant 1052/18. The work of M. Schwartz was supported in part by a German Israeli Project Cooperation (DIP) Grant under Grant PE2398/1-1. The work of M. Firer was supported in part by Fapesp, grant 13/25977-7 and CNPq 304046/2017-5.}
}

\maketitle

\begin{abstract}
  Motivated by an application to database linear querying, such as private information-retrieval protocols, we suggest a fundamental property of linear codes -- the generalized covering radius. The generalized covering-radius hierarchy of a linear code characterizes the trade-off between storage amount, latency, and access complexity, in such database systems. Several equivalent definitions are provided, showing this as a combinatorial, geometric, and algebraic notion. We derive bounds on the code parameters in relation with the generalized covering radii, study the effect of simple code operations, and describe a connection with generalized Hamming weights.
\end{abstract}

\begin{IEEEkeywords}
  Linear codes, covering radius, generalized Hamming weights, block metric
\end{IEEEkeywords}


\section{Introduction}
\label{sec:intro}

\IEEEPARstart{A}{ common} query type in database systems involves a linear combination of the database items with coefficients supplied by the user. As examples we mention partial-sum queries~\cite{chazelle1989computing}, and private information retrieval (PIR) protocols~\cite{chor1995private}. In essence, one can think of the database server as storing $m$ items, $x_1,\dots,x_m\in\F_{q^\ell}$. A user may query the contents of the database by providing $s_1,\dots,s_m\in\F_q$, and getting in response the linear combination $\sum_{i=1}^m s_i x_i$.

Various aspects of these systems are of interest and in need of optimization, such as the amount of storage at the server, and the required bandwidth for the querying protocol. One important such aspect is that of \emph{access complexity}, paralleling a similar concern studied in distributed storage systems \cite{GopHuaSimYek12,TamWanBru14}. In a straightforward implementation, the time required to access the elements of the database needed to compute the answer to a user query is directly proportional to the number of non-zero coefficients among $s_1,\dots,s_m$. This may prove to be a bottleneck, in particular since in schemes like PIR, the coefficients are random, and therefore a typical query would require the database server to access a fraction of $1-\frac{1}{q}$ of the items.

A trade-off between access complexity and storage amount was suggested for PIR in~\cite{ZhaYaaEtzSch19}, echoing a similar suggestion for databases made in~\cite{HoBruAgr98}. The suggestion calls for a carefully designed set of linear combinations to be pre-computed and stored by server. Instead of storing $\ox=(x_1,\dots,x_m)$ as is, the server stores $\oh_1\cdot\ox,\dots,\oh_n\cdot\ox$, where each $\oh_i\in\F_q^m$ describes a linear combination. Assume that the matrix $H$, whose columns are $\oh_1,\dots,\oh_n$, is a parity-check matrix for a code with covering radius $r$. Thus, when the user queries the database using $\os=(s_1,\dots,s_m)$, by the properties of the covering code, $\os$ may be computed using a linear combination of at most $r$ columns of $H$. Hence, at most $r$ pre-computed combinations that are stored in the database need to be accessed in order to provide the user with the requested linear combination. The trade-off between access complexity and storage amount follows, since instead of storing $m$ elements, the server now stores $n\geq m$ linear combinations, and so $n$ is lower bounded by the smallest possible length for a code with covering radius $r$ and redundancy $m$ over $\F_q$. These code parameters have been thoroughly studied and are well understood~\cite{Cohen}.

We now take access-complexity optimization one step further. The database server naturally receives a stream of queries, say $\os_1,\os_2,\dots$. Those may arrive from the same user, or from multiple distinct users. Instead of handling each of the queries separately, accessing $r$ pre-computed linear combinations for each query, the server may group together $t$ queries, $\os_1,\dots,\os_t$ and, hopefully, access fewer than $r\cdot t$ pre-computed linear combinations as it would in a naive implementation. Thus, both storage amount and latency are traded-off for a reduced access complexity.

The motivation mentioned above leads us to the following combinatorial problem: Design a set of vectors, $\oh_1,\dots,\oh_n\in\F_q^m$ (describing linear combinations to pre-compute), such that every $t$ vectors, $\os_1,\dots,\os_t\in\F_q^m$ (describing user queries), may be obtained by accessing at most $r$ of elements of $\oh_1,\dots,\oh_n$. When viewed as columns of a parity-check matrix for a code, this becomes a \emph{generalized covering radius} definition. It bears a resemblance to the generalized Hamming weight of codes, introduced by Wei~\cite{1991-Wei} to characterize the performance of linear codes over a wire-tap channel.

The goal of this paper is to study the generalized covering radius as a fundamental property of linear codes. Our main contributions are the following:
\begin{enumerate}
    \item 
    We discuss three definitions for the generalized covering radius of a code, highlighting the combinatorial, geometric, and algebraic properties of this concept, and showing them to be equivalent.
    \item
    We derive bounds that tie the various parameters of codes to the generalized covering radii. In particular, we prove an asymptotic upper bound on the minimum rate of binary codes with a prescribed second generalized covering radius, thus showing an improvement over the naive approach. The bound on the minimal rate is attained by almost all codes.
    \item
    We determine the effect simple code operations have on the generalized covering radii: code extension, puncturing, the $(u,u+v)$ construction, and direct sum.
    \item
    We discuss a connection between the generalized covering radii and the generalized Hamming weights of codes by showing that the latter is in fact a packing problem with some rank relaxation.
\end{enumerate}

The paper is organized as follows: Preliminaries and notations are presented in Section~\ref{sec:prelim}. We study various definitions of the generalized covering radius, and show them to be equivalent, in Section~\ref{sec:defs}. Section~\ref{sec:bounds} is devoted to the derivation of bounds on the generalized covering radii. Basic operations on codes are studied in Section~\ref{sec:ops}, and a relation with the generalized Hamming weights in Section~\ref{sec:genpack}. We conclude with a discussion of the results and some open questions in Section~\ref{sec:conc}.

\section{Preliminaries}
\label{sec:prelim}

For all $n\in\N$, we define $[n]\eqdef\set*{1,2,\dots,n}$. If $A$ is a finite set and $t\in\N$, we denote by $\binom{A}{t}$ the set of all subsets of $A$ of size exactly $t$. We use $\F_q$ to denote the finite field of size $q$, and denote $\F_q^*\eqdef \F_q\setminus\set*{0}$. Given a vector space $V$ over $\F_q$, we denote by $\sbinom{V}{t}$ the set of all vector subspaces of $V$ of dimension $t\in\N$. We use lower-letters, $v$, to denote scalars, overlined lower-case letters, $\ov$, to denote vectors, and either bold lower-case letters, $\bv$, or upper-case letter, $V$, to denote matrices. Whether vectors are row vectors or column vectors is deduced from context.

If $H$ is a matrix with $n$ columns, we denote by $\oh_i$ its $i$-th column. For $I=\set*{i_1,i_2,\dots,i_t}\in\binom{[n]}{t}$, we denote by $H_I$ the restriction of $H$ to the columns whose indices are in $I$, i.e., $H_I\eqdef ( \oh_{i_1}, \dots, \oh_{i_t})$. We shall also use $\spn*{H_I}$ to denote the linear space spanned by the columns of $H_I$, i.e.,
\[ \spn*{H_I}\eqdef \spn*{\oh_{i_1},\oh_{i_2},\dots,\oh_{i_t}}.\]

Given $\ov=(v_1,\dots,v_n)\in\F_q^n$, the support of $\ov$ is defined by
\[\supp(\ov)\eqdef \set*{i\in[n] ; v_i\neq 0}.\]
Whenever required, for a subset $V\subseteq\F_q^n$ we define
\[\supp(V)\eqdef \bigcup_{\ov\in V}\supp(\ov).\]
The Hamming weight of $\ov$ is then defined as $\wt(\ov)\eqdef\abs*{\supp(\ov)}$. If $\ov'\in\F_q^n$, then the Hamming distance between $\ov$ and $\ov'$ is given by $d(\ov,\ov')\eqdef \wt(\ov-\ov')$. We also extend the definition to the distance between a vector and a set, namely, for a set $C\subseteq\F_q^n$,
\[d(\ov,C)\eqdef \min\set*{ d(\ov,\oc) ; \oc\in C}.\]

Two shapes that will be useful to us are the ball and the cube. For a non-negative integer $r$, the Hamming ball of radius $r$ centered at $\ov\in\F_q^n$ is defined as
\[B_{r,n,q}(\ov)\eqdef\set*{\ov'\in\F_q^n ; d(\ov,\ov')\leq r}.\]
The cube with support $I\in\binom{[n]}{r}$ centered at $\ov\in\F_q^n$ is defined as
\[Q_{I,n,q}(\ov)\eqdef \set*{\ov'\in\F_q^n ; \supp(\ov'-\ov)\subseteq I}.\]
We shall omit the subscripts $n$ and $q$ whenever they may be inferred from the context. We observe that
\[ \bigcup_{I\in\binom{[n]}{r}}Q_I(\ov)=B_r(\ov).\]

\section{The Generalized Covering Radii}
\label{sec:defs}

We would now like to introduce the concept of generalized covering radius. We present several definitions, with varying approaches, be they combinatorial, algebraic, or geometric. We then show all of the definitions are in fact equivalent (at least, when linear codes are concerned).

Our first definition stems directly from the application outlined in the introduction -- database queries.

\begin{definition}
\label{def:rt1}
Let $C$ be an $[n,k]$ linear code over $\F_q$, given by an $(n-k)\times n$ parity-check matrix $H\in\F_q^{(n-k)\times n}$. For every $t\in\N$ we define the \emph{$t$-th generalized covering radius}, $R_t(C)$, to be the minimal integer $r\in\N$ such that for every set $S\in\binom{\F_q^{n-k}}{t}$ there exists $I\in\binom{[n]}{r}$ such that $S\subseteq \spn*{H_I}$. That is,
\[ R_t(C) \eqdef \max_{\substack{S\subseteq\F_q^{n-k}\\ \abs*{S}=t}}\min_{\substack{I\subseteq [n]\\ S\subseteq \spn*{H_I}}}\abs*{I}.\]
\end{definition}

While $R_t(C)$ certainly depends on the code $C$, for the sake of brevity we sometimes write $R_t$ when we can infer $C$ from the context. At first glance it seems as if $R_t$ does not only depend on $C$, but also on the choice of parity-check matrix $H$. However, the following lemma shows this is not the case.

\begin{lemma}
Let $R_t$ be the given by a full-rank matrix $H\in\F_q^{(n-k)\times n}$ as in Definition~\ref{def:rt1}. For any $A\in \gl(n-k,q)$ (the group of $(n-k)\times (n-k)$ invertible matrices with coefficients in $\F_q$), let $R'_t$ be the generalized covering radius, as in definition~\ref{def:rt1}, but using the matrix $AH$. Then $R_t=R'_t$.
\end{lemma}
\begin{IEEEproof}
Given $\os\in \F_q^{n-k}$, if $\os =\sum_{i\in I}\alpha_i\oh_i$, then, by linearity, we have that
\[A\os=A\sum_{i\in I}\alpha_i\oh_i=\sum_{i\in I}\alpha_iA\oh_i.\]
It follows that, given $S\subseteq\F_q^{n-k}$, if $S\subseteq\spn*{H_I}$, then $A(S)\subseteq\spn*{AH_I}$. Thus,
\[\min_{\substack{I\subseteq [n]\\ S\subseteq \spn*{H_I}}}\abs*{I}\geq\min_{\substack{I\subseteq [n]\\ A(S)\subseteq \spn*{AH_I}}}\abs*{I}.\]
Continuing with the same argument but using $A^{-1}$, we have
\[\min_{\substack{I\subseteq [n]\\ A(S)\subseteq \spn*{AH_I}}}\abs*{I}\geq \min_{\substack{I\subseteq [n]\\ A^{-1}A(S)\subseteq \spn*{A^{-1}AH_I}}}\abs*{I}=\min_{\substack{I\subseteq [n]\\ S\subseteq \spn*{H_I}}}\abs*{I}.\]
It then follows that
\[\min_{\substack{I\subseteq [n]\\ S\subseteq \spn*{H_I}}}\abs*{I}=\min_{\substack{I\subseteq [n]\\ A(S)\subseteq \spn*{AH_I}}}\abs*{I}.\]
As a consequence, if $S_0$ realizes the maximum condition,
\[R_t=\max_{\substack{S\subseteq\F_q^{n-k}\\ \abs*{S}=t}}\min_{\substack{I\subseteq [n]\\ S\subseteq \spn*{H_I}}}\abs*{I}=\min_{\substack{I\subseteq [n]\\ S_0\subseteq \spn*{H_I}}}\abs*{I} = \min_{\substack{I\subseteq [n]\\ A(S_0)\subseteq \spn*{AH_I}}}\abs*{I}.\]
It follows that $R_t\leq R'_t$. A symmetric argument, gives the reversed inequality, proving the desired claim.
\end{IEEEproof}

We observe, in Definition~\ref{def:rt1}, that requiring $S\subseteq\spn*{H_I}$ also ensures $\spn*{S}\subseteq\spn*{H_I}$. We therefore must have for all $t\in[n-k]$,
\begin{equation}
    \label{eq:lower}
    R_t\geq t.
\end{equation}
We also observe that $R_1$ is in fact the covering radius of the code $C$, and that the generalized covering radii are naturally monotone increasing, i.e.,
\begin{equation}
    \label{eq:monotone}
R_1\leq R_2 \leq \dots \leq R_{n-k}=n-k,
\end{equation}
as well as $R_t=n-k$ for all $t\geq n-k$. Thus, the values $R_1,\dots,R_{n-k}$ are called the \emph{generalized covering-radius hierarchy}. While being monotone increasing, we do note however, that the generalized covering radius $R_t$ is not necessarily strictly increasing in $t$, as the following example shows.

\begin{example}\label{exemp.Hamming}
Consider the binary Hamming code $C$, with parameters $[2^m-1,2^m-1-m,3]$. An $m\times (2^m-1)$ parity-check matrix $H$ for $C$ comprises of all binary vectors of length $m$ as columns, except for the all-zero column. One can easily check that $R_t(C)=t$ for all $t\in[m]$.

Assume $m\geq 2$. Now take $C'$ to be a $[2^m-2,2^m-2-m,3]$ code obtained from $C$ by shortening once. Thus, a parity-check matrix $H'$ for $C'$ is obtained by taking $H$  and deleting one of its columns; let us suppose that the shortening was done in the position corresponding to the all-ones column of $H$.  We now obviously have $R_1(C')=2$ since in order to cover $\set*{\oone}$ two columns of $H'$ are required. However, we also have $R_2(C')=2$ since any $2$-dimensional subspace of $\F_2^m$ has at least two nonzero (and hence linearly independent) vectors that appear as columns of $H'$.
\end{example}

Aiming for a geometric interpretation of the generalized covering radii, we provide two more equivalent definitions that are increasingly geometric in nature.

\begin{definition}
\label{def:rt2}
Let $C$ be an $[n,k]$ linear code over $\F_q$. Then for every $t\in\N$ we define the $t$-th generalized covering radius, $R_t(C)$, to be the minimal integer $r\in\N$ such that for every $\ov_1,\dots,\ov_t\in\F_q^n$, there exist codewords $\oc_1,\dots,\oc_t\in C$ and there exists $I\in\binom{[n]}{r}$, such that $\ov_i\in Q_I(\oc_i)$ for all $i\in[t]$.
\end{definition}

\begin{lemma}
\label{lem:samedef12}
Let $C$ be an $[n,k]$ linear code over $\F_q$. Then the values of $R_t$ from Definitions~\ref{def:rt1} and~\ref{def:rt2} are the same.
\end{lemma}
\begin{IEEEproof}
Fix a parity-check matrix $H$ for $C$ (with full rank). Denote the numbers from Definition \ref{def:rt1} and Definition \ref{def:rt2} by $R_t$ and $R_t'$, respectively. 

For the first direction, let $\ov_1,\dots,\ov_t\in \F_q^n$. Consider $\os_1,\dots,\os_t\in\F_q^{n-k}$ given by $\os_i=H \ov_i$ for all $i\in[t]$. By Definition~\ref{def:rt1} of $R_t$, there exists a set $\set*{i_1,\dots,i_{R_t}}=I\in \binom{[n]}{t}$ such that $\os_1,\dots,\os_n\in \spn*{H_I}$. That is, for each $\ell\in[t]$, there exist scalars $w_{\ell,1},\dots, w_{\ell,R_t}\in\F_q$ such that $\os_\ell=\sum_{j=1}^{R_t}w_{\ell,j}\oh_{i_j}.$ We define $\ow_{\ell}\in \F_q^n$ to be the vector containing $w_{\ell,1},\dots, w_{\ell,R_t}$ in the positions of $I$, and $0$ otherwise. Let $\oc_\ell\eqdef \ow_\ell-\ov_\ell$. We note that $\oc_\ell\in C$, as \[H\oc_\ell=H\ow_\ell-H\ov_\ell=\os_\ell-\os_\ell=\ozero.\]
On the other hand,$\supp(\oc_\ell-\ov_\ell)=\supp(\ow_\ell)\subseteq I$, and in particular $\ov_\ell\in Q_I(\oc_\ell)$. This shows that $R_t\geq R'_t$.

For the second direction of the proof, assume we have vectors $\os_1\dots,\os_t\in \F_q^{n-k}$. Since $H$ has full rank, there exist $\ov_1,\dots,\ov_t\in \F_q^n$ such that $H\ov_i=\os_i$ for all $i\in[t]$. From Definition~\ref{def:rt2} of $R_t'$, there exists a set $I\in \binom{[n]}{t}$ such that for all $i\in[t]$, $\supp(\ov_i-\oc_i)\subseteq I$. For each $i\in[t]$, we define $\ow_i\eqdef \ov_i-\oc_i$, and we have
\[ H \ow_i=H (\ov_i-\oc_i)=\os_i.\]
Since $\supp(\ow_i)\subseteq I$, for all $i\in[t]$, it follows that $s_1,\dots,s_t \in \spn*{H_I}$. This shows that $R_t'\geq R_t$.

Combining the two directions together we obtain that the values of $R_t$ from Definitions~\ref{def:rt1} and~\ref{def:rt2} are the same.
\end{IEEEproof}

We now move to a ``classical'' covering in the geometric sense. It involves a covering of a space with certain shapes. We shall require an extension of the cube to a $t$-cube. Given a non-negative integer $r$ and support $I\in\binom{[n]}{r}$, the $t$-cube centered at 
\[\bv=\begin{bmatrix} \ov_1 \\ \vdots \\ \ov_t\end{bmatrix}\in\F_q^{t\times n},\] 
is defined as
\begin{align*}
& Q^{(t)}_{I,n,q}(\bv)\eqdef \Bigg\{\bv'=\begin{bmatrix}\ov'_1 \\ \vdots \\ \ov'_t \end{bmatrix}\in \F_q^{t\times n} ~\Bigg|~ \\
&\qquad\qquad\qquad \forall i\in[t], \supp(\ov'_i-\ov_i)\in I\Bigg\}.
\end{align*}
This brings us to the definition of a $t$-ball centered at $\bv$ given by,
\[B^{(t)}_{r,n,q}(\bv)\eqdef \bigcup_{I\in\binom{[n]}{r}}Q^{(t)}_I(\bv),\]
where we say $r$ is the radius of the $t$-ball. Again, we shall omit the subscripts $n$ and $q$ whenever they may be inferred from the context. This is indeed a generalization of the ball since
\[ B^{(1)}_r(\bv)=B_r(\bv).\]
Thus, a superscript of $^{(1)}$ will generally be omitted unless a special need for emphasis arises.

In fact, the ball $B^{(t)}_r(\bv)$ realizes a ball in the natural sense, in the following metric we now define. The space we operate in is $\F_q^{t\times n}$. The \emph{$t$-weight} of a matrix $\bv\in\F_q^{t\times n}$, with row vectors denoted $\ov_i$, is defined as
\[\wt^{(t)}(\bv)\eqdef\abs*{\bigcup_{i\in[t]}\supp(\ov_i)}.\]
We now define the $t$-distance between $\bv,\bv'\in \F_q^{t\times n}$ as
\[ d^{(t)}(\bv,\bv')\eqdef \wt^{(t)}(\bv - \bv').\]
In particular, this also shows that $d^{(t)}$ is translation invariant, i.e., for all $\bv,\bv',\bv''\in \F_q^{t\times n}$,
\[ d^{(t)}(\bv+\bv'',\bv'+\bv'')=d^{(t)}(\bv,\bv').\]

It is easily seen now that the $t$-ball is in fact a ball in the metric induced by the $t$-distance, i.e.,
\[B^{(t)}_r(\bv)=\set*{\bv'\in \F_q^{t\times n} ; d^{(t)}(\bv,\bv')\leq r}.\]
We also note that $d^{(1)}$ is simply the Hamming distance function, hence our previous observation of a $1$-ball being a ball in the Hamming metric.

\begin{definition}
\label{def:rt3}
Let $C$ be an $[n,k]$ linear code over $\F_q$. Then for every $t\in\N$, we define the $t$-th generalized covering radius, $R_t$, to be the minimal integer $r$ such that $t$-balls centered at 
\[C^t \eqdef \set*{ \begin{bmatrix} \oc_1 \\ \vdots \\ \oc_t\end{bmatrix} ; \forall i\in[t], \oc_i\in C},\]
cover $\F_q^{t\times n}$, i.e.,
\[\bigcup_{\bc\in C^t}B^{(t)}_r(\bc) = \F_q^{t\times n}.\]
\end{definition}

\begin{lemma}
\label{lem:tcovering}
Let $C$ be an $[n,k]$ linear code over $\F_q$. Then the values of $R_t$ from Definitions~\ref{def:rt1},~\ref{def:rt2}, and~\ref{def:rt3}, are the same.
\end{lemma}
\begin{IEEEproof}
The proof is straightforward. We observe that for every $\ov_1,\dots,\ov_t\in\F_q^n$ there are $\oc_1,\dots,\oc_t\in C$ and a support $I\in\binom{[n]}{r}$ such that $\ov_i\in Q_I(\oc_i)$ for all $i\in[t]$ if and only if \[\begin{bmatrix} \ov_1 \\ \vdots \\ \ov_t \end{bmatrix}\in B_r^{(t)}\parenv*{\begin{bmatrix} \oc_1 \\ \vdots \\ \oc_t\end{bmatrix}}.\]
Thus, the minimal integer $r$ which defines $R_t$ is the same in Definitions~\ref{def:rt2} and~\ref{def:rt3}. By Lemma~\ref{lem:samedef12}, it is also the same as in Definition~\ref{def:rt1}.
\end{IEEEproof}

We would like to comment that if we denote the \emph{columns} of $\bv\in\F_q^{t\times n}$ by $\widehat{v}_1, \dots , \widehat{v}_n \in \F_{q}^t$, then
\[ \wt^{(t)}(\bv) = \abs*{ \set*{j \in [n] ; \widehat{v}_j \neq \ozero }}.
\]
This metric is known in the literature as the \emph{block metric} and it was introduced, independently, by Gabidulin \cite{Gabidulin} and Feng \cite{Feng}.

For our last approach, we make the obvious next step, resulting in an algebraic definition of the generalized covering radii. Assume $\bv\in\F_q^{t\times n}$ has rows $\ov_1,\dots,\ov_t\in\F_q^n$. Using the well known isomorphism $\F_q^t\cong \F_{q^t}$, we can then read each column of $\bv$ as a single element from $\F_{q^t}$. More precisely, fix a basis for $\F_{q^t}$ as a vector space over $\F_q$, say, $\beta_1,\dots,\beta_t\in \F_{q^t}$, and associate with $\bv$ above the vector
\begin{equation}
\label{eq:mapping}
\bv=\begin{bmatrix} \ov_1 \\ \vdots \\ \ov_t\end{bmatrix}\in \F_q^{t\times n}\quad \mapsto\quad  \Xi(\bv)\eqdef\sum_{i=1}^t \beta_i \ov_i \in\F_{q^t}^n.
\end{equation}
Note that this mapping is in fact a bijection. Under this mapping, a $t$-ball is mapped to a ball, namely,
\begin{equation}
\label{eq:ballmap}
\Xi(B^{(t)}_{r,n,q}(\bv)) = B^{(1)}_{r,n,q^t}(\Xi(\bv)),
\end{equation}
where we emphasize that the two balls are over different alphabets.

\begin{definition}
\label{def:rt6}
Let $C$ be an $[n,k]$ linear code over $\F_q$. Assume $G\in\F_q^{k\times n}$ is a generator matrix for $C$, namely, \[ C = \set*{ \ou G ; \ou\in\F_{q}^k}. \]
Let $t\in\N$, and let $C'$ be the linear code over $\F_{q^t}$ generated by the same matrix $G$, namely,
\[ C' = \set*{ \ou G ; \ou\in\F_{q^t}^k}. \]
Then we define the $t$-th generalized covering radius $R_t$ of $C$ as the covering radius of $C'$, namely,
\[ R_t(C) \eqdef R_1(C').\]
\end{definition}

\begin{lemma}
\label{lem:smallfield}
Let $C$ be an $[n,k]$ linear code over $\F_q$. Then the values of $R_t$ from Definitions~\ref{def:rt1},~\ref{def:rt2},~\ref{def:rt3}, and~\ref{def:rt6}, are the same.
\end{lemma}
\begin{IEEEproof}
Assume the notation of Definition~\ref{def:rt6}. Let $\bc\in C^t$, with rows $\oc_1,\dots,\oc_t\in C$, and let $\ou_i\in\F_q^k$ be such that $\oc_i=\ou_i G$. As in~\eqref{eq:mapping}, assume $\beta_1,\dots,\beta_t\in\F_{q^t}$ is a basis for $\F_{q^t}$ over $\F_q$. Then
\[ \Xi(\bc)=\sum_{i=1}^t \beta_i\oc_i=\parenv*{\sum_{i=1}^t \beta_i\ou_i}G.\]
Hence, $\Xi(\bc)\in C'$, where $C'$ is the code generated by $G$ over $\F_{q^t}$. A symmetric argument gives that $\Xi$ is in fact a bijection between $C^t$ and $C'$. The claim now follows from Definition~\ref{def:rt3} and Lemma~\ref{lem:tcovering}.
\end{IEEEproof}

As a final comment to this section, our original approach to generalize the covering radii of a code, introduced in Definition~\ref{def:rt1}, arises from the interest in querying databases by linear combinations (as, for example, used in PIR), and it uses the parity-check matrix of a code, hence it makes sense only for linear codes. This is not the case for the approach in Definition~\ref{def:rt3}, where $R_t$ is defined intrinsically as a metric invariant. This means that we can use this definition to generalize the covering radii for general (non-linear) codes.

\section{Bounds}
\label{sec:bounds}

A crucial part in our understanding of any figure of merit, is the limits of values it can take. Thus, we devote this section to the derivation of bounds on the generalized covering radii of codes. We put an emphasis on asymptotic bounds, that, given the normalized $t$-th covering radius, bound the best possible rate. We present a straightforward ball-covering argument for a lower bound. We then also present a trivial upper bound. Our main result is an asymptotic upper bound that improves upon the trivial one, and thus showing there is merit to the usage generalized covering radii to improve database querying, as described in Section~\ref{sec:intro}. Our upper bound is non-constructive, and uses the probabilistic method. It shall be made constructive (albeit, not useful) in Section~\ref{sec:ops}.

As is standard, we will require the size of a $t$-ball. Since the metrics involved are all translation invariant, the size of the ball does not depend on the choice of center. We therefore use \[V^{(t)}_{r,n,q}\eqdef
\abs*{B^{(t)}_{r,n,q}(\zero )}.\]
Thus, \eqref{eq:ballmap} gives the following immediate corollary.
\begin{corollary}
\label{cor:ballsize}
For all integers $n,t,r$ and a prime power $q$,
\[V^{(t)}_{r,n,q}=V_{r,n,q^t}=\sum_{i=0}^{r}\binom{n}{i}(q^t-1)^i.\]
\end{corollary}

We also recall the definition of the $q$-ary entropy function,
\[ H_q(x) = x\log_q (q-1) - x\log_q (x) - (1-x)\log_q (1-x).\]
Using Stirling's approximation, it is well known that
\[ V_{r,n,q} = \begin{cases}
q^{n H_q(r/n)-o(n)} & 0\leq \frac{r}{n} \leq 1-\frac{1}{q},\\
q^{n-o(n)}& 1-\frac{1}{q}<\frac{r}{n}\leq 1,
\end{cases}
\]
and thus,
\begin{equation}
    \label{eq:ballentropy}
V^{(t)}_{r,n,q}=V_{r,n,q^t}=
\begin{cases}
q^{tn H_{q^t}(r/n)-o(n)} & 0\leq \frac{r}{n} \leq 1-\frac{1}{q^t},\\
q^{tn-o(n)}& 1-\frac{1}{q^t}<\frac{r}{n}\leq 1.
\end{cases}
\end{equation}

Let $k_t(n,r,q)$ denote the smallest dimension of a linear code $C$ over $\F_q$ with length $n$ and $t$-covering radius $R_t(C)\leq r$. The following theorem was proved in~\cite{cohen1985good}.

\begin{theorem}[{\cite{cohen1985good}}]
\label{th:cohen}
For all $n,r\in\N$, and a prime power $q$,
\begin{align*}
n-\log_q V_{r,n,q} & \leq k_1(n,r,q) \\
& \leq n-\log_q V_{r,n,q}+2\log_2 n-\log_q n +O(1).
\end{align*}
\end{theorem}

It is convenient to study normalized parameters with respect to the length of the code. If $C$ is an $[n,k]$ linear code, we define its normalized parameters,
\begin{align*}
    \kappa&\eqdef \frac{k}{n}, & \rho_t&\eqdef\frac{R_t}{n}.
\end{align*}
Note that we use $\kappa$ for the rate of the code, and not $R$, to avoid confusion with the covering radius. For $t\in \N$ and a normalized covering radius $0\leq \rho\leq 1$, the minimal rate achieving $\rho$ is defined to be 
\[ \kappa_t(\rho,q)
\eqdef \liminf_{n\to \infty}\frac{k_t(n,\rho n,q)}{n}.\]
In this notation, Theorem~\ref{th:cohen} gives an asymptotically tight expression,
\begin{equation}
\label{eq:asympub}
\kappa_1(n,\rho) = \begin{cases}
1-H_q(\rho) & 0\leq \rho\leq 1-\frac{1}{q},\\
0 & 1-\frac{1}{q}<\rho\leq 1.
\end{cases}
\end{equation}

\subsection{General Bounds}

For a simple lower bound we use the ball-covering argument.

\begin{proposition}
\label{prop:AsymptoticLB}
For any $n,t\in \N$, prime power $q$, and $0\leq \rho \leq 1-\frac{1}{q^t}$, 
\[ \kappa_t(\rho,q)\geq 1-H_{q^t}(\rho).\]
\end{proposition}
\begin{IEEEproof}
Let $C$ be an $[n,k]$ code over $\F_q$ with $R_t(C)\leq \rho n$. For any $\bc\in C^t$ consider the $t$-ball of radius $R_t(C)$ centered at $\bc$, $B_{R_t(C)}^{(t)}(\bc)$. By Definition~\ref{def:rt3},
\[ \bigcup_{\bc\in C^t} B_{R_t(C)}^{(t)}(\bc)=\F_q^{t\times n}. \]
Thus, using Corollary~\ref{cor:ballsize},
\[ q^{kt}\cdot V^{(t)}_{R_t(C),n,q}=\abs*{C^t}\cdot V^{(t)}_{R_t(C),n,q}=\sum_{\bc\in C^t}\abs*{B_{R_t(C)}^{(t)}(\bc)}\geq q^{nt},\]
and therefore,
\[ \frac{k}{n}\geq 1-\frac{\log_{q^t}V^{(t)}_{R_t(C),n,q}}{n}.\]
Using~\eqref{eq:ballentropy} we get,
\[ \kappa \geq 1-  H_{q^t}\parenv*{\frac{R_t(C)}{n}}+o(1)\geq 1-H_{q^t}\parenv*{\rho}+o(1).\] 
This bound holds for an arbitrary $[n,k]$ code with $t$-covering radius at most $\rho n $. Therefore, we have  \[ \frac{k_t(n,\rho n,q)}{n}\geq 1-H_{q^t}\parenv*{\rho}+o(1), \]
and by taking $\liminf$ we conclude.
\end{IEEEproof}

For an upper bound, we first make the following observation.

\begin{proposition}
\label{prop:ttimes}
Let $C$ be an $[n,k]$ code over $\F_q$. Then for all $t\in\N$,
\[ R_t \leq t\cdot R_1.\]
\end{proposition}
\begin{IEEEproof}
Let $H$ be a parity-check matrix for $C$. By Definition~\ref{def:rt1}, given $S=\set*{\os_1,\dots,\os_t}\in\binom{\F_q^{n-k}}{t}$, there exist $I_i\in\binom{[n]}{R_1}$ such that $\os_i\in\spn{H_{I_i}}$, for all $i\in[t]$. Define $I\eqdef\bigcup_{i\in[t]}I_i$, then $S\subseteq\spn{H_I}$. It follows that
\[ R_t\leq \abs*{I} \leq \sum_{i=1}^{t}\abs{I_i}=t\cdot R_1.\]
\end{IEEEproof}

We can now give the following naive upper bound.

\begin{proposition}
\label{prop:NaiveUB}
For any $n,t\in \N$, $t\geq 2$, prime power $q$, and $0\leq \rho \leq 1$,
\[ \kappa_t(\rho,q)\leq 1-H_{q}\parenv*{\frac{\rho}{t}}.\]
\end{proposition}
\begin{IEEEproof}
By Proposition~\ref{prop:ttimes},
\[ \kappa_t(\rho,q)\leq \kappa_1\parenv*{\frac{\rho}{t},q}.
\]
We then combine \eqref{eq:asympub} with the fact that $t\geq2$ implies $\frac{\rho}{t}\leq 1-\frac{1}{q}$, to obtain the desired result.
\end{IEEEproof}

Proposition~\ref{prop:ttimes} is in fact a consequence of the following, more general, upper bound. This upper bound shows the generalized covering radii are sub-additive.

\begin{proposition}
\label{prop:subadd}
Let $C$ be an $[n,k]$ code over $\F_q$. Then for all $t_1,t_2\in\N$,
\[ R_{t_1+t_2} \leq R_{t_1}+R_{t_2}.\]
\end{proposition}
\begin{IEEEproof}
Let $H$ be a parity-check matrix for $C$. Given $S\in\binom{\F_q^{n-k}}{t_1+t_2}$, partition it arbitrarily to $S=S_1\cup S_2$, where $\abs{S_1}=t_1$ and $\abs{S_2}=t_2$. By Definition~\ref{def:rt1} there exist
\[I_1\in\binom{[n]}{R_{t_1}} \qquad\text{and}\qquad I_2\in\binom{[n]}{R_{t_2}},\]
such that $S_1\subseteq\spn{H_{I_1}}$, and $S_2\subseteq\spn{H_{I_2}}$. Define $I\eqdef I_1\cup I_2$, then $S=S_1\cup S_2\subseteq\spn{H_I}$. The claim now follows.
\end{IEEEproof}

\subsection{Upper Bounding the Binary Case with $t=2$}

The upper bound we now present improves upon the trivial one from Proposition~\ref{prop:NaiveUB}. Since it is significantly more complex, and has many moving parts, we focus on the binary case with $t=2$ only. We follow a similar strategy to the one employed by \cite[Theorem 12.3.5]{Cohen} for the covering radius, though major adjustments are required due to the more involved nature of this generalized problem. In essence, we show the existence of a covering code using the probabilistic method. The probability is nearly $1$, implying almost \emph{all} codes are at least as good as this bound. The main result is Theorem~\ref{th:mainupper}.

We outline the proof strategy to facilitate reading this section. We use the probabilistic method by choosing a random generator matrix for a code and bounding the probability that balls centered at the codewords indeed cover the entire space. To do so, we study the random variable that counts how many codewords cover a given point in space. To get a handle on this variable, we bound its expectation and variance.

We first recall the following useful lemma from~\cite[Lemma 1]{cohen1985good} concerning the average intersection of a set with its translations. Though originally proved for vectors, it also holds (with exactly the same proof) for matrices.

\begin{lemma}[\cite{cohen1985good}]
\label{lem:CosetAvg}
For any $S\subseteq \F_q^{t\times n}$,
\[ \frac{1}{q^{tn}}\sum_{\bv\in \F_q^{t\times n}}\abs*{S\cap (S+\bv)}=\frac{\abs*{S}^2}{q^{tn}}.\]
\end{lemma}

Let $k,n\in \N$ such that $n\geq 2$ and $t\leq k<n$. We consider the random matrix $G\in\F_2^{k\times n}$, with rows $\og_1,\dots, \og_k$ independently and uniformly drawn from $\F_2^n$. Let  $C$ be the random code with generator matrix $G$.

For a matrix $\bu\in \F_2^{t\times k}$, let $\bc_{\bu}\in C^t$ be defined by $\bc_{\bu} \eqdef \bu  G$. Clearly, \[ C^t=\set*{\bc_{\bu} ; \bu \in \F_2^{t\times k}}.\]
The next lemma shows a connection between the rank of $\bu$ and the statistical independence of the rows of $\bc_{\bu}$. We remark that the probability of $\bu$ being a full rank matrix goes to $1$ as $k\to\infty$. 

\begin{lemma}
\label{lem:uniform}
If $\bu\in\F_2^{t\times k}$ has full rank, then $\bc_{\bu}$ is uniformly distributed on $\F_2^{t\times n}$. In particular, the rows of $\bc_{\bu}$ are statistically independent. 
\end{lemma}
\begin{IEEEproof}
Consider the function $f_{\bu}:\F_2^{k\times n}\to \F_2^{t\times n}$ given by $f_{\bu}(A)=\bu A$. Since $\bu$ has full rank, $f_{\bu}$ is surjective and it is $2^{(k-t)n}$ to one. Hence, for any subset $S\subseteq \F_2^{t\times n}$, the size of the pre-image $f_{\bu}^{-1}(S)$ is $q^{(k-t)n}\abs{S}$. We recall that the generator matrix $G$ is uniformly distributed on $\F_2^{k\times n}$. Hence,
\begin{align*}
    \Prob\sparenv*{\bc_{\bu} \in S}&=\Prob\sparenv*{\bu G \in S}=\Prob\sparenv*{f_{\bu}(G) \in S}\\
    &=\Prob\sparenv*{G\in f_{\bu}^{-1}(S)}=\frac{\abs*{f_{\bu}^{-1}(S)}}{2^{kn}}\\
    &=\frac{2^{(k-t)n}|S|}{2^{kn}}=\frac{\abs*{S}}{2^{tn}}=\frac{\abs*{S}}{\abs*{\F_2^{t\times n}}}.
\end{align*}
This completes the proof.
\end{IEEEproof}

In preparation for bounding the variance of a certain random variable yet to be defined, we shall need to study the probability that pairs of codewords reside in the same ball. For $\bu_1,\bu_2\in \F_2^{2\times k}$, we consider the matrix  $\bc_{\bu_1,\bu_2}\in \F_2^{4\times k}$ defined by \[\ \bc_{\bu_1,\bu_2}\eqdef \begin{bmatrix}
\bc_{\bu_1}\\
\bc_{\bu_2}
\end{bmatrix}.\]
We first show that the probability the two codewords are contained in the same ball is maximized by the ball around $\zero$.

\begin{lemma}
\label{lem:Center0}
Let $1 \leq r \leq n-k$ be an integer and $\bu_1,\bu_2\in \F_2^{2\times k}$ with full rank, such that $\rank \begin{bsmatrix} \bu_1\\ \bu_2 \end{bsmatrix}=3$. Then for any $\bv\in \F_2^{2\times n}$ we have
\[ \Prob\sparenv*{\bc_{\bu_1,\bu_2}\in \parenv*{B_{r,n,2}^{(2)}(\bv)}^2 } \leq \Prob\sparenv*{\bc_{\bu_1,\bu_2}\in \parenv*{B_{r,n,2}^{(2)}(\zero)}^2} ,\]
where 
\[\parenv*{B_{r,n,2}^{(2)}(\bv)}^2=B_{r,n,2}^{(2)}(\bv)\times B_{r,n,2}^{(2)}(\bv)\subseteq \F_2^{4\times n}\]
is the Cartesian product of the ball $B_{r,n,2}^{(2)}(\bv)$ with itself. 
\end{lemma}
\begin{IEEEproof}
Let $\ou_1,\ou_2,\ou_3, \ou_4$ denote the rows of $\begin{bsmatrix} \bu_1\\ \bu_2 \end{bsmatrix}$  and $\bv=\begin{bsmatrix} \ov_1 \\ \ov_2\end{bsmatrix}\in \F_2^{2\times n}$. Without loss of generality, we assume that $\ou_1$, $\ou_2$, and $\ou_3$, are linearly independent. By this assumption, 
\[\ou_4=a_1\cdot \ou_1+a_2\cdot \ou_2+a_3\cdot \ou_3,\]
for some $a_1,a_2,a_3\in \F_2$. Let $\oc_1,\oc_2$ and $\oc_3,\oc_4$ be the rows of $\bc_{\bu_1}$ and $\bc_{\bu_2}$, respectively. We have 
\[  \begin{bmatrix} \oc_1 \\ \oc_2  \\ \oc_3 \\  \oc_4 \end{bmatrix}= \begin{bmatrix} \bc_{\bu_1} \\ \bc_{\bu_2}  \end{bmatrix}= \begin{bmatrix} \bu_1 \\ \bu_2  \end{bmatrix} \cdot G = \begin{bmatrix} \ou_1 \\ \ou_2  \\ \ou_3 \\ a_1\ou_1+a_2\ou_2+ a_3\ou_3 \end{bmatrix} \cdot G,  \] 
where $G$ is the random generator matrix of the code. Thus,
\[ \oc_4=a_1\cdot \oc_1+a_2\cdot \oc_2 +a_3\cdot \oc_3,\]
and by Lemma~\ref{lem:uniform}, $\begin{bsmatrix} \oc_1\\ \oc_2\\ \oc_3\end{bsmatrix}$ is uniformly distributed on $F_2^{3\times n}$.

We define 
\[ \begin{bmatrix} \tc_1 \\ \tc_2  \\ \tc_3 \\  \tc_4 \end{bmatrix}\eqdef \begin{bmatrix} \oc_1 \\ \oc_2  \\ \oc_3 \\  \oc_4 \end{bmatrix}-\begin{bmatrix} \ov_1 \\ \ov_2  \\ \ov_1 \\ \ov_2 \end{bmatrix}= \bc_{\bu_1,\bu_2} -  \begin{bmatrix} \bv\\ \bv\end{bmatrix}.  \] 
By the translations invariance of the metric $d^{(2)}$, 
\[  \bc_{\bu_1,\bu_2}\in \parenv*{B_{r,n,2}^{(2)}(\bv)}^2 \iff   \bc_{\bu_1,\bu_2} -  \begin{bmatrix} \bv\\ \bv\end{bmatrix} \in \parenv*{B_{r,n,2}^{(2)}(\zero )}^2.\] 

We note that the map $\psi:\F_2^{3\times n}\to \F_2^{3\times n}$ given by 
\[ \psi(\bz)=\bz- \begin{bmatrix} \ov_1 \\ \ov_2  \\ \ov_1 \end{bmatrix}\]
is a bijection, and therefore, 
\[\begin{bmatrix} \tc_1 \\ \tc_2  \\ \tc_3 \end{bmatrix}=\psi\parenv*{\begin{bmatrix} \oc_1 \\ \oc_2  \\ \oc_3 \end{bmatrix}}\]
is uniformly distributed on $\F_2^{3\times n}$ as well.

We divide our analysis into cases, depending on the value of $(a_1,a_2,a_3)$. Since $\rank(\bu_1)=\rank(\bu_2)=2$, and $\rank \begin{bsmatrix} \bu_1\\ \bu_2 \end{bsmatrix}=3$, the combinations $(a_1,a_2,a_3)=(0,0,0)$ and $(a_1,a_2,a_3)=(0,0,1)$ are impossible. 

\textbf{Case 1:} If $(a_1,a_2,a_3)=(1,1,1)$, a simple calculation shows that we have, 
\[\tc_4=\tc_1+\tc_2+\tc_3.\]
Thus, $\bc_{\bu_1,\bu_2}$ and $\bc_{\bu_1,\bu_2} -  \begin{bsmatrix} \bv\\ \bv\end{bsmatrix}$ have the same distribution, so
\begin{align*}
&\Prob\sparenv*{\bc_{\bu_1,\bu_2}\in \parenv*{B_{r,n,2}^{(2)}(\bv)}^2 } \\
&\qquad = \Prob\sparenv*{\bc_{\bu_1,\bu_2} -  \begin{bmatrix} \bv\\ \bv\end{bmatrix}\in \parenv*{B_{r,n,2}^{(2)}(\zero )}^2 }\\
&\qquad =\Prob\sparenv*{\bc_{\bu_1,\bu_2}\in \parenv*{B_{r,n,2}^{(2)}(\zero )}^2 }.
\end{align*}
    
\textbf{Case 2:} If $(a_1,a_2,a_3)=(1,1,0)$, a similar calculation as in the previous case shows that 
\[ \tc_4=\tc_1+\tc_2+\ov_1.\] 
For any $\oz\in \F_2^{n}$ we consider the set
\[ S_{\oz}\eqdef \set*{\begin{bmatrix}\ow_1 \\ \ow_2\\ \ow_3 \end{bmatrix}\in \F_2^{3\times n} ; \begin{bmatrix} \ow_1\\ \ow_2\\ \ow_3 \\ \ow_1+\ow_2+\oz\end{bmatrix}\in \parenv*{B_r^{(2)}(\zero )}^2 }.\]

Since $\begin{bsmatrix} \tc_1 \\ \tc_2  \\ \tc_3 \end{bsmatrix}$ and $\begin{bsmatrix} \oc_1 \\ \oc_2  \\ \oc_3 \end{bsmatrix}$ are uniformly distributed, to prove the theorem's claim is equivalent to showing that $\abs{S_{\ozero}}\geq \abs{S_{\ov_1}}$, which is also equivalent to $\abs{S_{\ozero}\setminus S_{\ov_1}}\geq \abs{S_{\ov_1}\setminus S_{\ozero}}$.

If $\ov_1=\ozero$, this condition is automatically satisfied. Otherwise, we will prove our claim by showing that for $\ov_1'$ obtained by zeroing one of the bits of $\ov_1$ we have $\abs{S_{\ov_1'}\setminus S_{\ov_1}}\geq \abs{S_{\ov_1}\setminus S_{\ov_1'}}$. Then, repeating this arguments and zeroing all the non-zero bits of $\ov_1$ we conclude the desired inequality.

Indeed, we find an injection $S_{\ov_1}\setminus S_{\ov_1'}\to S_{\ov_1'}\setminus S_{\ov_1}$. Let $i\in[n]$ be an index such that the $i$-th bit of $\ov_1$ is $1$. Denote by $\ove_i$ the $i$-th standard unit vector, and set $\ov_1'=\ov_1+\ove_i$. Let $\begin{bsmatrix} \ow_1\\ \ow_2\\ \ow_3\end{bsmatrix}\in S_{\ov_1}\setminus S_{\ov_1'}$. We have 
\[ \begin{bmatrix} \ow_1\\ \ow_2\end{bmatrix}\in B_r^{(2)}(\zero ), \quad \begin{bmatrix} \ow_3\\ \ow_1+\ow_2+\ov_1 \end{bmatrix}\in B_r^{(2)}(\zero ) \] 
and 
\[ \begin{bmatrix} \ow_3\\ \ow_1+ \ow_2 + \ov_1+ \ove_i \end{bmatrix}\notin B_r^{(2)}(\zero ). \] 
Thus,
\[
\supp(\ow_1+\ow_2 +\ov_1) \subsetneqq \supp(\ow_1+\ow_2 +\ov_1+\ove_i).
\]
Since the $i$-th bit of $\ov_1+\ove_i$ is $0$, it is only possible if the $i$-th bit of $\ow_1+\ow_2$ is $1$. Hence,
\begin{equation}
\label{eq:isupp}
i\in \supp\parenv*{\begin{bmatrix} \ow_1\\ \ow_2\end{bmatrix}}.
\end{equation}

We define 
\[\phi\begin{bmatrix} \ow_1\\ \ow_2\\ \ow_3\end{bmatrix} \eqdef \begin{bmatrix} \ow_1+ \ove_i\\ \ow_2\\ \ow_3\end{bmatrix}.\]
By~\eqref{eq:isupp} we have,
\[ \supp\parenv*{\phi\begin{bmatrix} \ow_1\\ \ow_2\\ \ow_3\end{bmatrix} }\subseteq \supp\parenv*{\begin{bmatrix} \ow_1\\ \ow_2\\ \ow_3\end{bmatrix} }.\]
Hence, 
\[\begin{bmatrix}\ow_1+\ove_i\\ \ow_2\end{bmatrix}\in B_r^{(2)}(\zero ). \]
Furthermore, $(\ow_1+\ow_2+\ove_i)+\ov_1'=\ow_1+\ow_2+\ov_1$, and so
\[ \begin{bmatrix} \ow_3\\ (\ow_1+\ow_2+\ove_i)+\ov_1'\end{bmatrix}=\begin{bmatrix}\ow_3\\ \ow_1+\ow_2+\ov_1\end{bmatrix}\in B_r^{(2)}(\zero ).\]
That is,
$\phi \begin{bsmatrix} \ow_1\\ \ow_2\\ \ow_3\end{bsmatrix}\in S_{\ov_1'}$. On the other hand, 
\[\begin{bmatrix} \ow_3\\ (\ow_1+\ow_2+\ove_i)+\ov_1\end{bmatrix}=\begin{bmatrix}\ow_3\\ \ow_1+\ow_2+\ov_1'\end{bmatrix}\notin B_r^{(2)}(\zero ). \]
Hence,  $\phi \begin{bsmatrix} \ow_1\\ \ow_2\\ \ow_3\end{bsmatrix}\notin S_{\ov_1}$. This shows that $\phi$ maps $S_{\ov_1}\setminus S_{\ov_1'}$ to $S_{\ov_1'}\setminus S_{\ov_1}$. Clearly $\phi$ is injective and it is the desired map.
     
\textbf{Case 3:} If $(a_1,a_2,a_3)=(1,0,1)$ we have $\oc_4=\oc_3+\oc_1$, or equivalently, $\oc_1=\oc_3+\oc_4$. This case is equivalent to the case where $(a_1,a_2,a_3)=(1,1,0)$  with $\oc_1,\oc_2$ and $\oc_3,\oc_4$ switching rolls. 

\textbf{Case 4:} If $(a_1,a_2,a_3)=(0,1,1)$ this is equivalent to the case where  $(a_1,a_2,a_3)=(1,1,0)$.
     
\textbf{Case 5:} If  $(a_1,a_2,a_3)=(0,1,0)$ we have
\[\tc_4=\tc_2. \]
Thus, $\bc_{\bu_1,\bu_2}$ and $\bc_{\bu_1,\bu_2} -  \begin{bsmatrix} \bv\\ \bv\end{bsmatrix}$ have the same distribution, and the case is completed as Case 1.

\textbf{Case 6:} If $(a_1,a_2,a_3)=(1,0,0)$ then we have 
\[\tc_4=\tc_1+(\ov_1+\ov_2). \]
Similarly to Case 2, where $(a_1,a_2,a_3)=(1,1,0)$, we show that $\abs{S_{\ozero}}\geq \abs{S_{\ov_1+\ov_2}}$. We use the same technique in order to show that we increase $\abs{S_{\ov_1+\ov_2}}$ when we flip a bit of $\ov_1+\ov_2$ from $1$ to $0$, and the same mapping $\phi$.
\end{IEEEproof}

For any $\bv\in \F_2^{2\times n}$ we define $X_{\bv}$ to be the number of codewords in $C^2$ that are generated from full-rank coefficients matrices, and that $r$-cover $\bv$. Formally, 
\[ X_{\bv}\eqdef \sum_{\substack{\bu \in \F_2^{2\times k}\\ \rank(\bu)=2}} \ind_{\set*{\bv \in B_r^{(2)}(\bc_{\bu})}}, \]
where $\ind_A$ is the indicator function of the event $A$. Clearly, $X_{\bv}$ depends on $n$, $k$, and $r$, although we omit them them from our notation. The random variable $X_{\bv}$ plays an important role in our main result, and we study its properties in preparation for the main theorem. We first bound the expectation of $X_{\bv}$.

\begin{lemma}
\label{lem:EBound}
For $0\leq\rho < \frac{3}{4}$, $n,k,r\in \N$, $3\leq k\leq n$, $r=\rho n$, and $\bv\in \F_2^{2\times n}$, 
\[V_{r,n,2}^{(2)}\cdot 2^{2k-1-2n} < \E[X_{\bv}] < V_{r,n,2}^{(2)}\cdot 2^{2k-2n}.\]
\end{lemma}
\begin{IEEEproof}
By Lemma~\ref{lem:uniform} for $\bu\in \F_2^{2\times k}$ with full rank, $\bc_{\bu}$ is uniformly distributed on $\F_2^{2\times n}$. Therefore,
\begin{align*}
    \E[X_{\bv}] &=\sum_{\substack{\bu \in \F_2^{2\times k}\\\rank(\bu)=2}}\E\sparenv*{ \ind_{\set*{\bv \in B_r^{(2)}(\bc_{\bu})}}}\\
    &=\sum_{\substack{\bu \in \F_2^{2\times k}\\\rank(\bu)=2}} \Prob\sparenv*{\bv \in B_r^{(2)}(\bc_{\bu})}\\
    &=\sum_{\substack{\bu \in \F_2^{2\times k}\\\rank(\bu)=2}} \frac{\abs*{B_r^{(2)}(\bc_{\bu})}}{2^{2n}}=(2^k-1)(2^k-2) \frac{V_{r,n,2}^{(2)}}{2^{2n}}.
\end{align*}
For $k\geq 3$ we have 
\[2^{2k-1}<(2^k-1)(2^k-2)<2^{2k},\]
which gives us the desired result.
\end{IEEEproof}

We consider the functions $f_1,f_2:[0,1]\to \R$ defined by
\begin{align*}
f_1(\rho)\eqdef \max_{\substack{0\leq\alpha\leq\rho \\ 0\leq\beta\leq \alpha \\ 0\leq \gamma\leq \rho-\alpha+\beta}}\Bigg( &  H_2(\alpha)+\alpha H_2\parenv*{\frac{\beta}{\alpha}}+2(\alpha-\beta)\\ &+(1-\alpha+\beta)H_2\parenv*{\frac{\gamma}{1-\alpha+\beta}}\Bigg),
\end{align*}
and 
\begin{align*}
     f_2(\rho)\eqdef \max_{\substack{0\leq \alpha\leq \rho \\ 0\leq \beta\leq \rho-\alpha}}\parenv*{  H_2(\alpha)+2(1-\alpha) H_2\parenv*{\frac{\beta}{1-\alpha}}+2\alpha)}.
\end{align*}
We then define
\[f(\rho)\eqdef \max\parenv*{f_1(\rho),f_2(\rho)},\]
which we will use in order to bound $\Var(X_{\bv})$.
\begin{lemma}
\label{lem:VarBound}
For $0\leq\rho< \frac{3}{4}$, $n,k,r\in \N$, $k\leq n$, $r=\rho n$, and $\bv\in \F_2^{2\times n}$, 
\[ \Var(X_{\bv})\leq 7\E[X_{\bv}]+2^{3(k-n)+n(f(\rho)+o(1))}\]
\end{lemma}
\begin{IEEEproof}
To simplify notation, we denote 
\[\eta_{\bu}\eqdef \ind_{\set*{\bv \in B_r^{(2)}(\bc_{\bu})}}.\]
We then calculate,
\begin{align}
    \Var(X_{\bv})&=\Var\parenv*{ \sum_{\substack{\bu \in \F_2^{2\times k}\\ \rank(\bu)=2}} \eta_{\bu}}\nonumber \\
    &= \sum_{\substack{\bu \in \F_2^{2\times k}\\ \rank(\bu)=2}} \Var\parenv*{ \eta_{\bu}} \nonumber \\
    & \quad\ + \sum_{\substack{\bu_1 \in \F_2^{2\times k}\\ \rank(\bu_1)=2}} \sum_{\substack{\bu_2 \in \F_2^{2\times k}\\ \rank(\bu_2)=2}} \Cov\parenv*{ \eta_{\bu_1}, \eta_{\bu_2}}. \label{eq:Var}
\end{align}
We separate the sums in \eqref{eq:Var} into four parts, and bound each one of them individually.

For a Bernoulli random variable $Z\sim \mathrm{Ber}(p)$ we have
\begin{equation}
\label{eq:vtoe}
\Var(Z)=\E[Z^2]-\E[Z]^2=p-p^2\leq p=\E[Z].
\end{equation}
Applying this bound to the first sum of \eqref{eq:Var}, we have
\begin{equation}
    \sum_{\substack{\bu \in \F_2^{2\times k}\\ \rank(\bu)=2}} \Var\parenv*{\eta_{\bu}}\leq \sum_{\substack{\bu \in \F_2^{2\times k}\\ \rank(\bu)=2}} \E\sparenv*{ \eta_{\bu}}=\E[X_{\bv}].
\end{equation}

We now consider the double sum in~\eqref{eq:Var}, which we separate into three parts, according to $\rank\parenv*{\begin{bsmatrix}  \bu_1\\ \bu_2\end{bsmatrix}}$.

If  $\rank\parenv*{\begin{bsmatrix}  \bu_1\\ \bu_2\end{bsmatrix}}=4$, by Lemma~\ref{lem:uniform}, $\bc_{\bu_1}$ and $\bc_{\bu_2}$ are statistically independent and  therefore so are $ \eta_{\bu_1} $ and  $\eta_{\bu_2}$. Thus, the covariance in that case is $0$.

If $\rank(\bu_1)=\rank(\bu_2)=\rank\parenv*{\begin{bsmatrix}  \bu_1\\ \bu_2\end{bsmatrix}}=2$,  by the Cauchy-Schwarz inequality, Lemma~\ref{lem:uniform} and \eqref{eq:vtoe} we have 
\begin{align*}
    \Cov\parenv*{ \eta_{\bu_1}, \eta_{\bu_2}}
    &\leq \sqrt{\Var\parenv*{ \eta_{\bu_1}} \Var\parenv*{ \eta_{\bu_2}}} \\
      &\leq \sqrt{\E\sparenv*{ \eta_{\bu_1}} \E\sparenv*{ \eta_{\bu_2}}} \\
     &=\sqrt{\Prob\sparenv*{\bv \in B_r^{(2)}(\bc_{\bu_1})}\Prob\sparenv*{\bv \in B_r^{(2)}(\bc_{\bu_2})}}\\
      &=\sqrt{\Prob\sparenv*{\bc_{\bu_1} \in B_r^{(2)}(\bv)}\Prob\sparenv*{\bc_{\bu_2} \in B_r^{(2)}(\bv)}}\\
     &=\frac{\abs*{B_r^{(2)}(\bv)}}{2^{2n}}=\E\sparenv*{ \eta_{\bu_1}}.
\end{align*}

Since $\rank(\bu_1)=\rank(\bu_2)$, we have that $\bu_2=A\cdot \bu_1$ where $A\in\gl(2,2)$ is a $2\times 2$ invertible matrix over $\F_2$.
Summation over all pairs $\bu_1,\bu_2$ such that $\rank(\bu_1)=\rank(\bu_2)=\rank\parenv*{\begin{bsmatrix}  \bu_1\\ \bu_2\end{bsmatrix}}=2$ gives 
\begin{align*}
&\sum_{\substack{\bu_1 \in \F_2^{2\times k}\\ \rank(\bu_1)=2}}\sum_{A\in\gl(2,2)}\Cov\parenv*{ \eta_{\bu_1}, \eta_{A\bu_1}}\\
&\qquad \leq \sum_{\substack{\bu_1 \in \F_2^{2\times k}\\ \rank(\bu_1)=2}}\sum_{A\in\gl(2,2)} \E\sparenv*{ \eta_{\bu_1}}\\
&\qquad =\sum_{A\in\gl(2,2)}\sum_{\substack{\bu_1 \in \F_2^{2\times k}\\ \rank(\bu_1)=2}}\E\sparenv*{ \eta_{\bu_1}}\\
&\qquad =\sum_{A\in\gl(2,2)}\E[X_{\bv}]=6\E[X_{\bv}],
\end{align*}
where the last equality follows since there are exactly six $2\times 2$ invertible matrices over $\F_2$.

We are left with the case of $\rank\parenv*{\begin{bsmatrix} \bu_1 \\ \bu_2\end{bsmatrix}}=3$. We start by bounding $\Cov(\eta_{\bu_1},\eta_{\bu_2})$, and then evaluate this bound, dividing our analysis into three cases according to the linear dependence structure of $\bu_1$ and $\bu_2$. 

By Lemma~\ref{lem:Center0},
\begin{align*}
    \Cov(\eta_{\bu_1},\eta_{\bu_2})&=\E[\eta_{\bu_1}\eta_{\bu_2}]-\E[\eta_{\bu_1}]\E[\eta_{\bu_2}]\\
    &\leq \E[\eta_{\bu_1}\eta_{\bu_2}]=\Prob\sparenv*{\bc_{\bu_1},\bc_{\bu_1}\in B_r^{(2)}(\bv)}\\
    &\leq \Prob\sparenv*{\bc_{\bu_1},\bc_{\bu_1}\in B_r^{(2)}(\zero )}.
\end{align*}
As before, let $\ou_1,\ou_2,\ou_3, \ou_4$ denote the rows of $\begin{bsmatrix} \bu_1\\ \bu_2 \end{bsmatrix}$. Without loss of generality, assume $\ou_1,\ou_2,\ou_3$ are linearly independent and $\ou_4\in\spn*{\ou_1,\ou_2,\ou_3}$, that is, 
\[ \ou_4=a_1\cdot\ou_1+a_2\cdot\ou_2+a_3\cdot\ou_3,\]
for some $a_1,a_2,a_3\in \F_2$. Thus, $\begin{bsmatrix} \ou_1 \\ \ou_2 \\ \ou_3\end{bsmatrix}G$ is uniformly distributed on $\F_2^{3\times n}$ and therefore $\Prob\sparenv*{\bc_{\bu_1},\bc_{\bu_2}\in B_r^{(2)}(\zero )}$ is proportional to the number of triples $\ow_1,\ow_2,\ow_3\in \F_2^n$ such that 
\[ \begin{bmatrix} \ow_1 \\ \ow_2 \end{bmatrix}, \begin{bmatrix} \ow_3 \\ a_1\cdot \ow_1+a_2\cdot \ow_2+a_3\cdot \ow_3 \end{bmatrix}\in B_r^{(2)}(\zero ).\]

We enumerate the number of such triples in every dependence structure, which we denote by $N(a_1,a_2,a_3)$.  Since $\rank(\bu_2)=2$ the combinations $a_1=a_2=a_3=0$ and $a_1=a_2=0$, $a_3=1$ are impossible. Thus we have six cases, and we will show that they can be reduced to three cases. 

\underline{Case 1 - $(a_1,a_2,a_3)=(1,1,0)$}: If $(a_1,a_2,a_3)=(1,1,0)$, then $\ow_4=\ow_1+\ow_2$. Hence, the number of triples in this case is given by
\[ N(1,1,0)= \sum_{i=0}^r \binom{n}{i}\sum_{j=0}^i \binom{i}{j}2^{i-j}\sum_{\ell=0}^{r-i+j}\binom{n-i+j}{\ell}2^{i-j}.\]
Here, the integer $i$ runs over all possible values for the size of $\supp\parenv*{\begin{bsmatrix} \ow_1\\ \ow_2 \end{bsmatrix}}$, namely, between $0$ and $r$, and $\binom{n}{i}$ counts the number of ways to choose this support. The integer $j$ runs over all possible values for the number of overlapping $1$'s  between $\ow_1$ and $\ow_2$, $\binom{i}{j}$ counts the number of ways to choose these overlapping positions, and $2^{i-j}$ counts the number of ways to distribute the remaining $1$'s between $\ow_1$ and $\ow_2$. The integer $\ell$ runs over all possible values of the number of non-overlapping $1$'s between $\ow_3$ and $\ow_1+\ow_2$, and $\binom{n-i+j}{\ell}$ counts the number of ways to choose those non-overlapping $1$'s in the remaining $n-(i-j)$ coordinates. Finally, $2^{i-j}$ counts the number of ways to choose overlapping $1$'s from $\supp(\ow_1+\ow_2)$ to $\ow_3$.
\underline{Case 2 - $(a_1,a_2,a_3)=(1,1,1)$}: By similar calculations as in the first case we obtain,
\[ N(1,1,1)= N(1,1,0).\]
\underline{Case 3 - $(a_1,a_2,a_3)=(1,0,0)$}: In this case
\[ N(1,0,0)= \sum_{i=0}^r \binom{n}{i}\parenv*{\sum_{j=0}^{r-i} \binom{n-i}{j}2^{i}}^2.\]
The other three cases can be reduced to one of the previous one. The case where $(a_1,a_2,a_3)=(0,1,0)$ is trivially equivalent to the case of $(1,0,0)$. If $(a_1,a_2,a_3)=(1,0,1)$ we have $\ow_4=\ow_1+\ow_3$,  and therefore $\ow_1=\ow_3+\ow_4$. Thus, this case is equivalent to the case of $(1,1,0)$ with $\ow_1,\ow_2$ and $\ow_3,\ow_4$ switching parts. Similarly the  case where $(a_1,a_2,a_3)=(0,1,1)$ is equivalent to the case of $(1,1,0)$. 

We recall that $r=\rho n$. Fix some $0\leq i \leq r$, $0\leq j\leq i$ and $0\leq \ell\leq \rho-i+j $. We denote 
\[ i = \alpha n \quad j=\beta n \quad \ell=\gamma n.\]
The constraints on $i,j$ and $\ell$ impose 
\[       0\leq \alpha \leq \rho, \quad 0\leq \beta\leq \alpha, \quad 0\leq \gamma\leq \rho-\alpha+\beta.\]
Using the well known identity $\binom{n}{\alpha  n}=2^{n(H_2(\alpha)+o(1))}$ we have
\begin{align*}
&\binom{n}{i} \binom{i}{j}2^{i-j}\binom{n-i+j}{\ell}2^{i-j}\\
&\qquad =2^{n\parenv*{H_2(\alpha)+\alpha H_2\parenv*{\frac{\beta}{\alpha}}+2(\alpha-\beta) +(1-\alpha+\beta)H_2\parenv*{\frac{\gamma}{1-\alpha+\beta}}+o(1)}}
\\&\qquad \leq 2^{n\parenv*{f_1(\rho) +o(1)}}. 
\end{align*}
and therefore 
\begin{align*}
&\sum_{i=0}^r \binom{n}{i}\sum_{j=0}^i \binom{i}{j}2^{i-j}\sum_{\ell=0}^{r-i+j}\binom{n-i+j}{\ell}2^{i-j}
\\&\qquad \leq n^3 2^{n\parenv*{f_1(\rho) +o(1)}}=2^{n\parenv*{f_1(\rho) +o(1)}}. 
\end{align*}

In a similar fashion we obtain 
\begin{align*}
    \sum_{i=0}^r \binom{n}{i}\parenv*{\sum_{j=0}^{r-i} \binom{n-i}{j}2^{i}}^2\leq 2^{n(f_2(\rho)+o(1))}
\end{align*}
Combining the bounds we obtain
\begin{align*}
    \Cov(\eta_{\bu_1},\eta_{\bu_2})&\leq \Prob\sparenv*{\bc_{\bu_1},\bc_{\bu_1}\in B_r^{(2)}(\zero )}\\
    &\leq \frac{2^{n(\max (f_1(\rho),f_2(\rho))+o(1))}}{2^{3n}}\\
    &=2^{n(f(\rho)-3+o(1))}.
\end{align*}
Summing over all $\bu_1,\bu_2$ such that $\rank\parenv*{\begin{bsmatrix}\bu_1 \\ \bu_2 \end{bsmatrix}}=3$ gives 
\begin{align*}
    &\sum_{\rank\parenv*{\begin{bsmatrix}\bu_1 \\ \bu_2 \end{bsmatrix}}=3} \Cov(\eta_{\bu_1},\eta_{\bu_2})\\
    &\qquad \leq  \sum_{\rank\parenv*{\begin{bsmatrix}\bu_1 \\ \bu_2 \end{bsmatrix}}=3}2^{n(f(\rho)-3+o(1))}\\
    &\qquad \leq\parenv*{2^k-1}\parenv*{2^k-2}\parenv*{2^k-4}2^32^{n(f(\rho)-3+o(1))}\\
    &\qquad \leq 2^{n(f(\rho)-3+o(1))+3k}.
\end{align*}
Summing the upper bounds on all of the parts in the sum  \eqref{eq:Var} we obtain the desired bound and complete the proof.
\end{IEEEproof} 

\begin{figure}[t]
\centering
\begin{overpic}[scale=0.7]
    {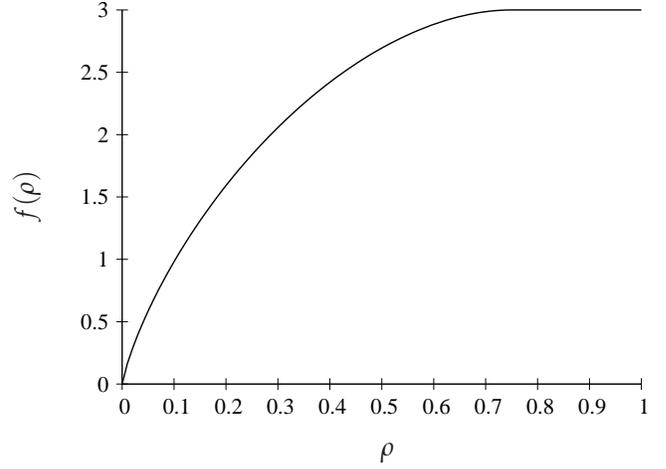}
    \put(0,35){\begin{turn}{90}$f(\rho)$\end{turn}}
    \put(55,0){$\rho$}
\end{overpic}
\caption{
The function $f(\rho)$ from Lemma~\ref{lem:VarBound}.
}
\label{fig:frho}
\end{figure}

The functions $f_1(\rho)$ and $f_2(\rho)$ are given by maximizing the two functions,
\begin{align*}
f'_1(\alpha,\beta,\gamma)&\eqdef \Bigg( H_2(\alpha)+\alpha H_2\parenv*{\frac{\beta}{\alpha}}+2(\alpha-\beta)\\ &\qquad +(1-\alpha+\beta)H_2\parenv*{\frac{\gamma}{1-\alpha+\beta}}\Bigg),\\
f'_2(\alpha,\beta)&\eqdef \parenv*{  H_2(\alpha)+2(1-\alpha) H_2\parenv*{\frac{\beta}{1-\alpha}}+2\alpha)}.
\end{align*}
We observe that the parameter $\rho$ only controls the maximization domain. Using standard analysis techniques we can find the exact expression for $f(\rho)$. Let us denote 
\[ s(\rho)\eqdef \frac{1}{10}\parenv*{1+8\rho -\sqrt{1+16\rho-16\rho^2}}. \] 
For $0\leq \rho<\frac{3}{4}$, $f'_1$ is maximized at the point
\[ \parenv*{\alpha_1^{\max}(\rho),   \beta_1^{\max}(\rho), \gamma_1^{\max}(\rho)}=\parenv*{\rho,\rho-s(\rho),s(\rho)},\] 
and for $\frac{3}{4}\leq \rho \leq 1$ it is maximized at 
\[  \parenv*{\alpha_1^{\max}(\rho),   \beta_1^{\max}(\rho), \gamma_1^{\max}(\rho)}=\parenv*{\frac{3}{4},\frac{1}{4},\frac{1}{4}}.\]
Similarly, for $0\leq \rho<\frac{3}{4}$, $f'_2$ is maximized at the point
\[  \parenv*{\alpha_2^{\max}(\rho),   \beta_2^{\max}(\rho)}=\parenv*{s(\rho),\rho-s(\rho)},\] 
and for $\frac{3}{4}\leq \rho \leq 1$ it is maximized at 
\[ \parenv*{\alpha_2^{\max}(\rho),   \beta_2^{\max}(\rho)}=\parenv*{\frac{1}{2},\frac{1}{4}}. \]
Curiously, $f'_1$ and $f'_2$ take the same value at their maximum points for any $0\leq \rho\leq 1$, and therefore 
\begin{align*}
    f(\rho)&=f'_1\parenv*{\alpha_1^{\max}(\rho),   \beta_1^{\max}(\rho), \gamma_1^{\max}(\rho)}\\
    &=f'_2\parenv*{\alpha_2^{\max}(\rho),   \beta_2^{\max}(\rho)}\\
    &=\begin{cases}
        H_2(s(\rho))+2s(\rho) & \\
        \quad +2(1-s(\rho))H_2\parenv*{\frac{\rho-s(\rho)}{1-s(\rho)}} & 0\leq \rho< \frac{3}{4}, \\
        3 & \frac{3}{4}\leq\rho\leq 1.
    \end{cases}
\end{align*}
The function $f(\rho)$ is shown in Figure~\ref{fig:frho}.

Having bounded the expectation and variance of $X_{\bv}$, we are ready to prove (under certain conditions) that the probability that all $X_{\bv}$ are positive, namely, the entire space is covered.

\begin{proposition}
\label{prop:CodesCover}
Let $0\leq \rho<\frac{3}{4}$, and let $(k_n)_{n=1}^\infty$ be a sequence of integers such that 
\[k_n>n-\log_4\parenv*{V_{\rho n ,n,2}^ {(2)}}+\log_2(n)\]
and 
\[\limsup_{n\to\infty}\max_{\ov\in \F_2^{2\times n}} \frac{\log_2(\Var(X_{\bv}))}{\log_2(\E[X_{\bv}])}< s<2,\]
where $X_{\bv}$ is defined with respect to $n$, $k_n$, and $r_n=\rho n$.

Let $C'_n$  be the random code  with a uniformly distributed $k_n'\times n$ random generator matrix $G'$, where $k_n'=k_n+2\ceil*{\log_2(n)}+2$. Then 
\[R_2(C'_n)\leq r_n = \rho n\]
with probability that tends to $1$ as $n\to\infty$.   
\end{proposition}
\begin{IEEEproof}
 Let $G$ be the matrix obtained by taking the first $k_n$ rows of $G'$, and $\og_1,\dots, \og_m$ be the remaining rows, $m=2\ceil*{\log_2(n)}+2$. We fix some $\varepsilon>0$ and $\bv\in F_2^{2\times n}$. Consider the random code $C_n$ with random generator matrix $G$. By Chebyshev's inequality 
\[ \Prob\sparenv*{|X_{\bv}-\E[X_{\bv}]|\geq 2^\varepsilon\cdot \E[X_{\bv}]^{s/2}}\leq \frac{\Var(X_{\bv})}{2^{2\varepsilon}\E[X_{\bv}]^s}.\]
By assumption, for sufficiently large $n$, $\Var(X_{\bv})< \E[X_{\bv}]^s$, and therefore 
\begin{align}
    \label{eq:prob1}
    \Prob\sparenv*{|X_{\bv}-\E[X_{\bv}]|\geq 2^\varepsilon\cdot \E[X_{\bv}]^{s/2}} <2^{-2\varepsilon}.
\end{align} 
We denote $\beta\eqdef\E[X_{\bv}]$ and define $\beta(\varepsilon)\eqdef \beta- 2^{\varepsilon}\beta^{s/2}$. From \eqref{eq:prob1} it follows that for sufficiently large $n$,
\[ \Prob\sparenv*{X_{\bv}\leq \beta(\varepsilon)} <2^{-2\varepsilon}. \] 

Define the set $Q_0$ to be the set of elements in $\F_2^{2\times n}$ that are $r_n$-covered by at most $\beta(\varepsilon)$ codewords from the set $\set*{\bc_{\bu} ; \bu\in \F_2^{2\times k}, \rank(\bu)=2}\subseteq C_n^2$. Formally,
\[  Q_0\eqdef \set*{\bv\in \F_2^{2\times n} ; X_{\bv}\leq \beta(\varepsilon)}. \]
Let  
\[q_0 \eqdef \frac{\abs*{Q_0}}{2^{2n}}\]
denote the proportion of $Q_0$ inside $\F_2^{2\times n}$. 
We have
\[ q_0=\frac{1}{2^{2n}} \sum_{\bv\in \F_2^{2\times n }} \ind_{\set*{X_{\bv}\leq \beta(\varepsilon)}},  \]
and therefore, for sufficiently large $n$,
\begin{equation}
\label{eq:q0expecatation}
     \E[q_0]=\Prob[X_{\bv}\leq \beta(\varepsilon)]< 2^{-2\varepsilon}.
 \end{equation}
 By Markov's inequality and \eqref{eq:q0expecatation} we have 
 \begin{equation}
     \label{eq:qEpsilonProb}
     \Prob\sparenv*{q_0\geq 2^{-\varepsilon}}
     \leq \frac{\E[q_0]}{2^{-\varepsilon}}
     < 2^{-\varepsilon}.
 \end{equation}

For a set $V\subseteq \F_2^{2\times n}$ we define $Q(V)$ to be the set of $r_n$-remote points from $V$. That is, 
\[ Q(V)\eqdef \F_2^{2\times n}\setminus \bigcup_{\bv\in V}B_{r_n}^{(2)}(\bv).\]
 Clearly, if $\beta(\varepsilon)\geq 0$, we have 
\begin{align}
\label{eq:QepsilonQ}
    Q(C_n)\subseteq Q_0.
\end{align}
We will later choose $\varepsilon$ such that  $\beta(\varepsilon)\geq 0$.

For arbitrary $\ox,\oy\in \F_2^n$ we consider the linear code generated by adding $\ox$ and $\oy$, which is $C_n+\spn*{\ox,\oy}$. From the definition of $Q$ we have,
 \begin{align*}
     Q\parenv*{(C_n+ \spn*{\ox,\oy})^2}&\subseteq  Q\parenv*{C_n^2\cup \parenv*{C_n^2+\begin{bmatrix} \ox \\ \oy \end{bmatrix}}}\\
     &= Q\parenv*{C_n^2}\cap Q\parenv*{C_n^2+\begin{bmatrix} \ox \\ \oy \end{bmatrix}}\\
     &= Q\parenv*{C_n^2}\cap \parenv*{Q\parenv*{C_n^2}+\begin{bmatrix} \ox \\ \oy \end{bmatrix}}.
 \end{align*}
 where the last equality follows from the invariance  of $d^{(2)}$ under translations. Hence, we have 
 \begin{equation*}
     \abs*{ Q\parenv*{(C_n+ \spn*{\ox,\oy})^2}}\leq \abs*{Q\parenv*{C_n^2}\cap \parenv*{Q\parenv*{C_n^2}+\begin{bmatrix} \ox \\ \oy \end{bmatrix}}}.
 \end{equation*}
 Combining this with \eqref{eq:QepsilonQ} and Lemma~\ref{lem:CosetAvg} we have 
 \begin{align}
  &\frac{1}{2^{2n}}\sum_{\ox,\oy\in \F_2^n}  \abs*{Q\parenv*{(C_n+ \spn*{\ox,\oy})^2}}\nonumber\\
  &\qquad \leq    \frac{1}{2^{2n}}\sum_{\ox,\oy\in \F_2^n}  \abs*{Q\parenv*{C_n^2}\cap \parenv*{Q\parenv*{C_n^2}+\begin{bmatrix} \ox \\ \oy \end{bmatrix}}}\nonumber\\
  &\qquad =\frac{1}{2^{2n}} \cdot \frac{\abs*{Q\parenv*{C_n^2}}^2}{2^{2n}}\leq  \frac{1}{2^{2n}} \cdot \frac{\abs*{Q_0}^2}{2^{2n}}=q_0^2. \label{eq:qExpctation}
 \end{align}

Recall the first $k_n$ rows of $G'$ make up $G$, and the remaining rows are denoted by $\og_1,\dots,\og_m$. For $1\leq \ell\leq \frac{m}{2}$ we denote
\[q_\ell\eqdef \frac{\abs*{Q\parenv*{(C_n+\spn*{\og_1,\dots,\og_{2\ell}})^2}}}{2^{2n}}.\]
Since the rows of $G$ and the remaining rows $\og_1,\dots,\og_m$ are independent and uniformly distributed, \eqref{eq:qExpctation} implies that
\begin{equation}
\label{eq:qIneq1}
    \E\cond*{q_1 ; q_0}\leq q_0^2,
\end{equation}
and by a similar argument, 
\begin{equation}
\label{eq:qIneq2}
    \E\cond*{q_\ell ; q_{\ell-1}}\leq q_{\ell-1}^2.
\end{equation}

We fix some $\lambda>0$ and bound $\Prob\sparenv*{q_1\leq2^{\lambda-2\varepsilon}}$ from below. 
By Markov's inequality, the law of total probability, and \eqref{eq:qEpsilonProb}, we have
\begin{align*}
    &\Prob\sparenv*{q_1\leq 2^{\lambda-2\varepsilon}}\\
    &\qquad \geq \Prob\cond*{q_1\leq 2^{\lambda-2\varepsilon} ; q_0\leq 2^{-\varepsilon} }\cdot \Prob\sparenv*{q_0\leq 2^{-\varepsilon}}  \\
    &\qquad \geq \parenv*{1-\Prob\cond*{q_1> 2^{\lambda-2\varepsilon} ; q_0\leq 2^{-\varepsilon} }}\cdot \parenv*{1-2^{-\varepsilon}}\\
    &\qquad\geq \parenv*{1-\frac{\E\cond*{q_1 ; q_0\leq 2^{-\varepsilon} }}{2^{\lambda-2\varepsilon}}}\cdot \parenv*{1-2^{-\varepsilon}}.
\end{align*}
By the law of total expectation and \eqref{eq:qIneq1} we obtain a bound on $\E\cond*{q_1 ; q_0\leq 2^{-\varepsilon} }$,
\begin{align*}
    &\E\cond*{q_1 ; q_0 \leq 2^{-\varepsilon} }\\
    &\qquad=\E_{q_0}\sparenv*{\E\cond*{q_1 ; q_0, q_0\leq 2^{-\varepsilon} }}  \\ 
    &\qquad=\sum_{a\leq 2^{-\varepsilon}}\E\cond*{q_1 ; q_0=a, q_0\leq 2^{-\varepsilon} }\Prob\cond*{q_0=a ; q_0\leq 2^{-\varepsilon}}\\
    &\qquad=\sum_{a\leq 2^{-\varepsilon}}\E\cond*{q_1 ; q_0=a }\Prob\cond*{q_0=a ; q_0\leq 2^{-\varepsilon}}\\
    &\qquad\leq \sum_{a\leq 2^{-\varepsilon}} a^2 \cdot \Prob\cond*{q_0=a ; q_0\leq 2^{-\varepsilon}}\leq 2^{-2\varepsilon},
\end{align*}
where summation over $a\leq 2^{-\varepsilon}$ is valid since our distribution is discrete with finite support. Altogether, 
\begin{equation*}
     \Prob\sparenv*{q_1\leq 2^{\lambda-2\varepsilon}}\geq \parenv*{1-2^{-\lambda}}\parenv*{1-2^{-\varepsilon}}.
\end{equation*}
Repeating the same arguments inductively using \eqref{eq:qIneq2}, we obtain
\begin{equation}
\label{eq:qProb}
    \Prob\sparenv*{q_\ell\leq 2^{2^\ell(\lambda-\varepsilon)-\lambda}}\geq \parenv*{1-2^{-\lambda}}^\ell \parenv*{1-2^{-\varepsilon}}
\end{equation}
for all $1\leq  \ell \leq \frac{m}{2}$. 

We now set  
\[\varepsilon=2\log_2(\log_2 n),\]
and recall the assumption,
\[k_n>n-\log_4\parenv*{V_{\rho n ,n,2}^ {(2)}}+\log_2(n).\]
By Lemma~\ref{lem:EBound} we have 
\begin{align*}
    \beta&=\E[X_{\bv}]>V_{\rho n ,n,2}^{(2)} \cdot 2^{2 k_n-1-2n}\\
    &>V_{\rho n ,n,2}^{(2)}2^{2\parenv*{n-\log_4\parenv*{V_{\rho n ,n,2}^ {(2)}}+\log_2(n)}-1-2n}=\frac{n^2}{2}.
 \end{align*} 
We also recall that in the beginning of our analysis, we assumed that $\varepsilon$ is chosen such that $\beta(\varepsilon)\geq 0$ (for sufficiently large $n$). Indeed, since $\beta>\frac{n^2}{2}$ and $s<2$,
\begin{align*}
    \beta(\varepsilon)&=\beta-2^{\varepsilon}\beta^{\frac{s}{2}}=\beta-2^{2\log_2(\log_2n))}\beta^{\frac{s}{2}}\\
    &=\beta-(\log_2n)^2\beta^{\frac{s}{2}}\xrightarrow[n\to\infty]{}\infty.
\end{align*} 

We set $\lambda=\varepsilon-1$. For sufficiently large $n$, $\lambda>0$ and 
\begin{align*}
    2^{2^{m/2}(\lambda-\varepsilon)-\lambda}&<2^{-2^{m/2}}=2^{-2^{\ceil*{\log_2(n)}+1}}< 2^{-2n}.
\end{align*} Hence, the event $\set*{q_{\frac{m}{2}}<2^{2^{m/2}(\lambda-\varepsilon)-\lambda}}$ implies the event that $\parenv*{C_n+\spn*{\og_1,\dots,\og_m}}^2=C_n'$  has covering radius $R_2(C_n')\leq \rho n$, because $q_{\frac{m}{2}}<2^{-2n}$ implies
\[\abs*{Q(C'_n)}=\abs*{Q\parenv*{(C_n+\spn*{\og_1,\dots,\og_{m}})^2}}<1,\]
but since this quantity is a non-negative integer, it must be $0$. In total, 
\begin{align*}
    &\Prob\sparenv*{R_2(C_n') \leq \rho n }\geq \parenv*{1-2^{-\varepsilon}}\parenv*{1-2^{-\lambda}}^{m/2}\\
    &\qquad = \parenv*{1-2^{-\varepsilon}}\parenv*{1-2^{-\varepsilon+1}}^{\ceil*{\log_2 n} +1}\\
    &\qquad =\parenv*{1- (\log_2 n)^{-2}} \parenv*{1-2 (\log_2 n)^{-2}}^{\ceil*{\log_2 n} +1}\\
    &\qquad=O\parenv*{1-\frac{1}{(\log_2(n))^2}}\xrightarrow[n\to\infty]{}1.
\end{align*}
This completes the proof.
\end{IEEEproof}

Ultimately, we are interested in $\kappa_t(\rho,q)$, which is the asymptotic minimal rate required for covering $(\F_q^n)^t$ (that we identify with $\F_q^{t\times n}$) by $t$-balls of radius $\rho n$. We now prove our main result for this section.

\begin{theorem}
\label{th:mainupper}
For any $0< \rho\leq 1$, 
\[ \kappa_2(\rho,2)\leq \begin{cases}
1-(4H_4(\rho)-f(\rho)) & 0\leq \rho< \frac{3}{4}, \\
0 & \frac{3}{4}\leq \rho\leq 1.
\end{cases}\]
\end{theorem}

\begin{IEEEproof}
Let us first consider $0\leq \rho < \frac{3}{4}$. Fix some $\varepsilon>0$.  For any $n\in \N$ we look for $k_n$ such that, if $n$ is sufficiently large,
\[ \frac{\log_2\Var(X_{\bv})}{\log_2 \E[X_{\bv}]}<2-\frac{\varepsilon}{2}\]
for all $\bv\in\F_2^{2\times n}$.
By Lemma~\ref{lem:VarBound},
\[ \Var(X_{\bv})\leq 7\E[X_{\bv}]+2^{3(k_n-n)+n(f(\rho)+o(1))}. \]
By Lemma~\ref{lem:EBound} we have
\[ \E[X_{\bv}]\geq  2^{2k_n-1-2n}V^{(2)}_{\rho n,n,2}=2^{2k_n-1-2n+2n(H_4(\rho)+o(1))},\] 
and therefore, 
\[ \frac{\log_2\Var(X_{\bv})}{\log_2 \E[X_{\bv}]}\leq \frac{3(k_n-n)+n(f(\rho)+o(1))}{2k_n-1-2n+2n(H_4(\rho)+o(1))}. \]
Let $k^*_n$ be the solution of the equation
\[ \frac{3(k^*_n-n)+nf(\rho)}{2k^*_n-1-2n+2n H_4(\rho)}= 2-\varepsilon, \] 
namely,
\[k^*_n=\frac{2-\varepsilon +n(1+f(\rho)-4H_4(\rho)+2\varepsilon(H_4(\rho)-1))}{1-2\varepsilon}.\]
Define $k_n\eqdef \floor*{k^*_n}$. Since $\frac{k_n}{k^*_n}\xrightarrow[n\to \infty]{} 1$, for sufficiently large $n$,
\begin{align*}
    \frac{\log_2\Var(X_{\bv})}{\log_2 \E[X_{\bv}]}&\leq \frac{3(k_n-n)+n(f(\rho)+o(1))}{2k_n-1-2n+2n(H_4(\rho)+o(1))}\\
    &\leq (2-\varepsilon)+\frac{\varepsilon}{2}=2-\frac{\varepsilon}{2}.
\end{align*} 

We note that 
\[\lim_{\varepsilon\to 0}k^*_n = 2+n(1-4H_4(\rho)+f(\rho)).\]
By standard analysis techniques, for all $0\leq \rho<\frac{3}{4}$,
\[ 1-4H_4(\rho)+f(\rho)>1-H_4(\rho).\]
Thus, for a sufficiently small $\varepsilon$ we can choose a sufficiently large $n$, such that
\[k_n> n-\log_4(V^{(2)}_{\rho n,n, 2})+\log_2(n)\]
and 
\[\frac{\log_2 \Var(X_{\bv})}{\log_2 \E[X_{\bv}]}\leq 2-\frac{\varepsilon}{2}.\]

By Proposition~\ref{prop:CodesCover}, there exists a sequence of codes $(C'_n)_{n=N}^\infty $ such that $C'_n$ is an $[n,k'_n]$ code where 
\begin{align*}
k'_n&=k_n+2\ceil*{\log_2(n)}+2,\\
R_2(C_n)&\leq \rho n.
\end{align*}
Thus, for sufficiently large $n$, 
\[ k_2(n,\rho n,2)\leq k_n+2\ceil*{\log_2(n)}+2,\]
and therefore,
\begin{align*}
    \kappa_2(\rho,2)&=\limsup_{n\to\infty}\frac{ k_2(n,\rho n,2)}{n}\\
     &\leq \lim_{n\to\infty}\frac{k_n+2\ceil*{\log_2(n)}+2}{n}\\
     &=\frac{1+f(\rho)-4H_4(\rho)+2\varepsilon(H_4(\rho)-1))}{1-2\varepsilon}.
\end{align*}
Taking $\varepsilon\to 0$ we obtain 
\[ \kappa_2(\rho,2)\leq 1+f(\rho)-4H_4(\rho),\]
as desired.

By its definition, $k_t(\rho,q)$ is a decreasing monotonic function in $\rho$. It is easy to verify that $f(\rho)$ tends to $3$ when $\rho$ tends to $\frac{3}{4}$ from the left. Thus, for  $\frac{3}{4}\leq \rho \leq 1$,
\begin{align*}
    0\leq k_2(\rho,2)&\leq \lim_{\rho' \to \parenv*{\frac{3}{4}}_-} \kappa_2(\rho',2)\\
    &\leq\lim_{\rho' \to \parenv*{\frac{3}{4}}_-} 1-4H_4(\rho')+f(\rho')=0.
\end{align*}
\end{IEEEproof}

\begin{figure*}[t]
\centering
\begin{overpic}[scale=1.25]
    {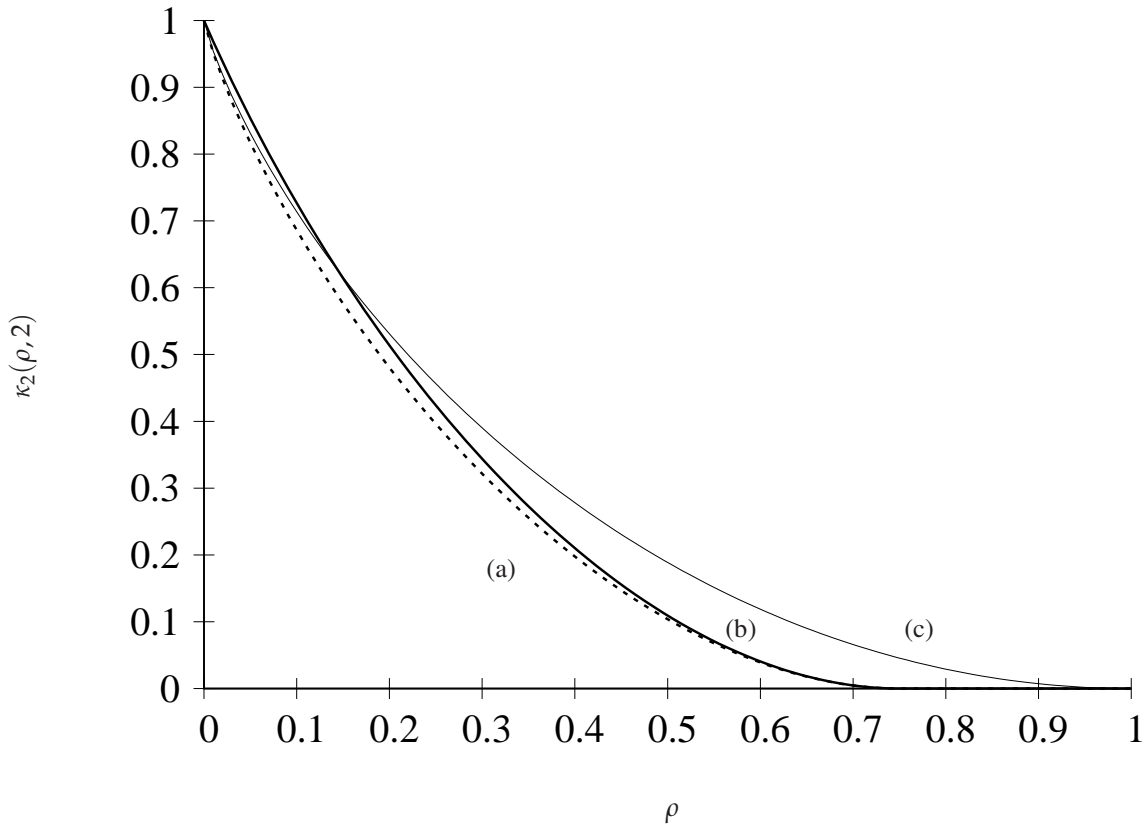}
    \put(0,35){\begin{turn}{90}$\kappa_2(\rho,2)$\end{turn}}
    \put(55,0){$\rho$}
    \put(40,20){(a)}
    \put(60,15){(b)}
    \put(75,15){(c)}
\end{overpic}
\caption{
A comparison of the bounds on $\kappa_2(\rho,2)$: (a) the ball-covering lower bound, (b) the upper bound of Theorem~\ref{th:mainupper}, and (c) the naive upper bound of Proposition~\ref{prop:NaiveUB}.
}
\label{fig:comp}
\end{figure*}

A comparison of the various asymptotic bounds is shown in Figure~\ref{fig:comp}. It is interesting to note that the upper bound of Theorem~\ref{th:mainupper} matches the lower ball-covering bound at $\rho=\frac{3}{4}$, particularly so because the function $f(\rho)$ is defined by the binary entropy function, and not the quaternary entropy function. We also note that the naive upper bound of Proposition~\ref{prop:NaiveUB} is better than the upper bound of Theorem~\ref{th:mainupper} for $\rho\lesssim 0.145$.

\section{Simple Code Operations}
\label{sec:ops}

Some code operations are very common. Among these we can find code extension, code puncturing, the $(u,u+v)$ construction, and direct sum. In this section we show the effect these operations have on the generalized covering radii mimics their effect on the (regular) covering radius. We use the direct product to turn the non-constructive upper bound of Theorem~\ref{th:mainupper} to an explicit construction, albeit, not a very useful one.

Given a code $C \subseteq \F_{q}^n$, let  
\[C^* \eqdef \set*{(c_1,\ldots,c_{n-1}) ; (c_1,\ldots,c_{n-1},c_n) \in C },\]
be the \emph{punctured code}, and  
\[\oC \eqdef \set*{(c_1,\ldots,c_n,-\sum_{i=1}^n c_i) ; (c_1,\ldots,c_n) \in C },\]
be the \emph{extended code}. Even though puncturing is defined as the removal of the last coordinate, the following results apply to the removal of any single coordinate.

By~\cite[Theorem 3.1.1, p. 62]{Cohen},  $R_1(C^*)$ equals $R_1(C)$ or $R_1(C) - 1$ and $R_1(\overline{C})$  equals $R_1(C)$ or $R_1(C) + 1$. The same result holds for the generalized covering radii.

\begin{proposition}
\label{prop:puncturing}
Let  $C$ be an $[n,k]$ linear code. Then for any $t\in\N$,
\begin{enumerate}
\item
$R_t(C^*)$ equals $R_t(C)$ or $R_t(C) - 1$;
\item	
$R_t(\overline{C})$ equals $R_t(C)$ or $R_t(C) + 1$.
\end{enumerate}
\end{proposition}
\begin{IEEEproof}
Let $G\in\F_q^{k\times n}$ be a generator matrix for $C$, and let $D\subseteq\F_{q^t}^{n}$ be the code generated by $G$ over $\F_{q^t}$. The code $C^*$ over $\F_q$ is then generated by $G^*$ which is the matrix obtained from $G$ by removing the last column. Denote by $D^*\subseteq\F_{q^t}^{n-1}$ the code generated by $G^*$ over $\F_{q^t}$. Obviously, $D^*$ is also obtained by puncturing $D$. By Definition~\ref{def:rt6}, and~\cite[Theorem 3.1.1, p. 62]{Cohen},
\begin{align*}
R_t(C^*)=R_1(D^*)&\in\set*{R_1(D),R_1(D)-1}\\
&=\set*{R_t(C),R_t(C)-1}.
\end{align*}

Using a similar approach we prove the case of code extension. We construct the following $k\times(n+1)$ matrix
\[ \oG\eqdef [ G,-G\cdot\oone].\]
Obviously the code generated over $\F_q$ by $\oG$ is $\oC$. Using $G$ and $\oG$ over $\F_{q^t}$ we get the codes $D$ and $\oD$, respectively. Again, by Definition~\ref{def:rt6}, and~\cite[Theorem 3.1.1, p. 62]{Cohen},
\begin{align*}
R_t(\oC)=R_1(\oD)&\in\set*{R_1(D),R_1(D)+1}\\
&=\set*{R_t(C),R_t(C)+1}.
\end{align*}
\end{IEEEproof}

Assume $C_1$ and $C_2$ are $[n,k_1]$ and $[n,k_2]$ codes, respectively.
The $(u,u+v)$ construction uses $C_1$ and $C_2$ to produce a code
\[ C=\set*{(\ou,\ou+\ov) ; \ou\in C_1,\ov\in C_2},\]
and by~\cite[Theorem 3.4.1, p.~66]{Cohen}, its covering radius is upper bounded by $R_1(C)\leq R_1(C_1)+R_1(C_2)$.

\begin{proposition}
\label{prop:uuplusv}
Let $C_i$ be an $[n,k_i]$ code over $\F_q$, $i=1,2$, and let $C$ be the code constructed from $C_1$ and $C_2$ using the $(u,u+v)$ construction. Then for any $t\in\N$,
\[ R_t(C)\leq R_t(C_1)+R_t(C_2).\]
\end{proposition}
\begin{IEEEproof}
If $G_i\in\F_q^{k_i\times n}$ is a generator matrix for $C_i$, $i=1,2$, then it is easy to see that
\[ G\eqdef \begin{bmatrix} G_1 & G_1 \\ 0 & G_2 \end{bmatrix},\]
is a generator matrix for $C$. The rest of the proof follows that of Proposition~\ref{prop:puncturing} by considering the code generated by $G$ over $\F_{q^t}$.
\end{IEEEproof}

We now look at the direct sum. Given an $[n_1,k_1]$ code $C_1$, and an $[n_2,k_2]$ code $C_2$, both over $\F_q$, the direct sum is defined as
\[ C_1\oplus C_2 \eqdef \set*{ (\oc_1,\oc_2); \oc_1\in C_1, \oc_2\in C_2},\]
which is an $[n_1+n_2,k_1+k_2]$ code over $\F_q$. It is well known~\cite[Theorem 3.2.1, p. 63]{Cohen} that
\[ R_1(C_1\oplus C_2)=R_1(C_1)+R_1(C_2).\]

\begin{proposition}
\label{prop:directproduct}
Let $C_i$ be an $[n_i,k_i]$ code over $\F_q$, for $i=1,2$. Then for any $t\in\N$,
\[R_t(C_1 \oplus C_2) = R_t(C_1) + R_t(C_2).\]
\end{proposition}
\begin{IEEEproof}
If $G_i\in\F_q^{k_i\times n_i}$ is a generator matrix for $C_i$, $i=1,2$, then it is easy to see that
\[G\eqdef \begin{bmatrix} G_1 & 0 \\ 0 & G_2\end{bmatrix},\]
is a generator matrix for $C_1\oplus C_2$. The rest of the proof follows that of Proposition~\ref{prop:puncturing} by considering the code generated by $G$ over $\F_{q^t}$.
\end{IEEEproof}

\begin{remark} The upper bound presented in Theorem~\ref{th:mainupper} is proved by showing the existence of a sequence of codes in a non-constructive way. We use Proposition~\ref{prop:directproduct} in order to find an explicit construction for a code attaining the bound of Theorem~\ref{th:mainupper}.

We fix $0\leq \rho \leq 1$. In the proof of Theorem~\ref{th:mainupper}, we find a sequence of covering codes $(C_n)_n$, where $C_n$ is an $[n,k_n]$ code with covering radius at most $\rho n$ and
\[\lim_{n\to\infty}\frac{k_n}{n}=1-4H_4(\rho)+f(\rho).\]
The existence of such a sequence of codes is guaranteed by Proposition~\ref{prop:CodesCover}, where it is proved that, when we randomly choose a $k_n\times n$ generator matrix, the probability to get a code with those properties is lower bounded by $O\parenv*{1-(\log_2(n))^{-2}}$.

We consider the $[n 2^{n\cdot k_n},k_n 2^{n\cdot k_n}]$ code generated by the direct product
\[ \widetilde{C}_n\eqdef \bigoplus_{G\in \F_2^{k_n\times n}} C_G, \] where $C_G$ is the code with generator matrix $G$. We note that the rate of $\widetilde{C}_n$ is upper bounded by $\frac{k_n}{n}$. By our probabilistic argument, the normalized covering radius of $\widetilde{C}_n$ is upper bounded by $\rho\cdot (1-O(\log_2(n)^{-2}))+1\cdot O(\log_2(n)^{-2})$, which tends to $\rho$ as $n\to\infty$. 

This technique of explicitly constructing codes by direct-summing codes is well known and has been used many times in order to make probabilistic proofs constructive, e.g., \cite{potukuchi2020improved,blinovskii1990covering}. The disadvantage of this technique is the enormous block length of the resulting code. In our construction, in order to ensure a normalized covering radius at most $\rho+\varepsilon$ the required block length is $\Omega\parenv{2^{2^{1/\varepsilon}+1/\sqrt{\varepsilon}}}$.
\end{remark}

\section{The Generalized Packing Radii}
\label{sec:genpack}

Given an $[n,k]$ linear code $C$ over $\F_q$, the generalized Hamming weight of the code, $d_t$, $t\in\N$, is defined as the minimal support size containing a linear subcode of $C$ of dimension $t$, i.e.,
\[ d_t \eqdef \min_{C'\in\sbinom{C}{t}} \abs*{\supp(C')}.\]
In particular, $d_1$ is the usual minimum distance of $C$.

Generalized Hamming weights were introduced by Wei in 1991 \cite{1991-Wei}, as a figure of merit to analyze the security performance of a code on a wire-tap channel. Wei proved that the weight hierarchy is strictly increasing and proved the duality theorem, relating the weight hierarchy of a code and its dual. A stronger duality theorem, namely a generalization of MacWilliams identity for the generalized Hamming weight distribution of a code and its dual was provided in \cite{KLOVE1992}. The weight hierarchy of a code was computed for many families of codes \cite{1991-Wei,1992-Feng,1998-Heijnen-Pellikaan} and bounds are produced in \cite{1992-Helleseth,ashikhmin1999new,cohen1994upper}. Natural generalizations of MDS codes are presented in \cite{dougherty2010higher}. The generalized weights were also defined in other metric instances, such as the rank metric \cite{oggier, Kurihara}. In a very interesting approach to generalized weights, considering a representation of linear codes as a set of points in a projective space, Tsfasman and Vladut \cite{tsfasman1995geometric} transform the generalized weights from a metric problem into a combinatorial-incidence problem. Forney showed in \cite{Forney1994} deep connections between the  generalized Hamming weight hierarchy of a linear code and the complexity of its minimal trellis diagram  and  an initial attempt to bound the error probability of a code (in the erasure channel) using the generalized weights was done in \cite{Didier2006, Lemes}.

In the following we shall require the size $\floor*{(d_t-1)/2}$. To simplify the presentation we define for all $t\in\N$,
\[\delta_t \eqdef \floor*{\frac{d_t-1}{2}}.\]
We also define the set
\[ \cL^{(t)}(\F_q^n)\eqdef \set*{\bv\in \F_q^{t\times n} ; \rank(\bv)=t}.\]

\begin{lemma}
\label{lem:tpacking}
Let $C$ be an $[n,k]$ linear code over $\F_q$. Then for every $t\in[k]$, $\delta_t$ is the largest integer satisfying that for all $\bc,\bc'\in C^t$ such that $\bc-\bc'\in\cL^{(t)}(\F_q^n)$,
\[ B^{(t)}_{\delta_t}(\bc)\cap B^{(t)}_{\delta_t}(\bc')=\emptyset.\]
\end{lemma}
\begin{IEEEproof}
First, for $0\leq r\leq\delta_t$, assume to the contrary that there exist $\bc,\bc'\in C^t$, $\bc-\bc'\in\cL^{(t)}(\F_q^n)$, and that
\[ B^{(t)}_{r}(\bc)\cap B^{(t)}_{r}(\bc')\neq\emptyset.\]
Let $\bv$ be in that intersection. Thus, there exists $I,I'\subseteq[n]$, with $\abs*{I},\abs*{I'}\leq r\leq \delta_t$, such that
\[ \bv\in Q^{(t)}_I(\bc) \qquad\text{and}\qquad \bv\in Q^{(t)}_{I'}(\bc').\]
But then
\[ \bc-\bc'\in Q^{(t)}_{I\cup I'}(\zero ).\]
Since $\bc-\bc'\in\cL^{(t)}(\F_q^n)$, the row space of $\bc-\bc'$ is a $t$-dimensional subcode of $C$ supported by $\abs*{I\cup I'}\leq 2r\leq 2\delta_t < d_t$ coordinates, which is a contradiction to the definition of $d_t$.

For the second direction, assume $r>\delta_t$. By the definition of $\delta_t$, there exists a subcode $C'=\spn*{\oc_1,\dots,\oc_t}\subseteq C$ of dimension $t$ and support $I$ of size $\abs*{I}=\delta_t$. Set $\bc\in C^t$ to be the matrix whose rows are $\oc_1,\dots,\oc_t$, and arbitrarily choose $I_1,I_2\in\binom{[n]}{r}$ such that $I\subseteq I_1\cup I_2$. We construct, for all $i\in[t]$, a vector $\ov_i$ that agrees with $\oc_i$ in the coordinates $I_1$, and is $0$ elsewhere. Set $\bv\in\F_q^{t\times n}$ to be the matrix whose rows are $\ov_1,\dots,\ov_t$, and observe that
\[\bv\in Q^{(t)}_{I_1}(\zero ) \qquad\text{and}\qquad \bv\in Q^{(t)}_{I_2}(\bc).\]
Hence
\[ B^{(t)}_{r}(\zero )\cap B^{(t)}_{r}(\bc)\neq\emptyset,\]
so $\delta_t$ is the maximal integer with the desired property.
\end{IEEEproof}

We observe that for $t=1$, Lemma~\ref{lem:tpacking} becomes the standard packing of Hamming error balls induced by the code $C$, and $\delta_1$ is the packing radius of the code, and hence, $\delta_1\leq R_1$. It is therefore tempting to conjecture that $\delta_t\leq R_t$ for all $t\in[\min\set{k,n-k}]$. However, Lemma~\ref{lem:tpacking} does not describe a packing of $t$-balls, when $t\geq 2$, since these may intersect if the difference between their centers is not of full rank.

\section{Conclusion}
\label{sec:conc}

We proposed a fundamental property of linear codes -- the generalized covering-radius hierarchy. It characterizes the trade-off between storage amount, latency, and access complexity in databases queried by linear combinations, as is the case, for example, in PIR schemes. We showed three equivalent definitions for these radii, highlighting their combinatorial, geometric, and algebraic aspects. We derived bounds on the code parameters in relation with the generalized covering radii, and studied the effect simple code operations have on them. Finally, we described a connection between the generalized covering-radius hierarchy and the generalized Hamming weight hierarchy.

While the study of the generalized covering-radius hierarchy has its own independent intellectual merit, let us also place the bound of Theorem~\ref{th:mainupper} back in the context of PIR schemes. Consider the binary case, and assume we allow a latency of $t=2$, namely, the server waits until two queries arrive and then handles them both. Further assume, that to handle the two queries we allow the server to access at most $\frac{1}{2}$ of its storage. Stated alternatively, the average access per query is a $\frac{1}{4}$ of the storage. By Theorem~\ref{th:mainupper}, since $\kappa_2(\frac{1}{2},2)\approx 0.11$, there exists a code allowing $89\%$ of the server storage for user information and only $11\%$ for overhead. A naive approach, using $\kappa_1(\frac{1}{4},2)\approx 0.19$, implies the storage may contain only $81\%$ user information and $19\%$ overhead.

Many other open problems remain, and we mention but a few. First, extending Theorem~\ref{th:mainupper} to address non-binary generalized covering radii for all $t$ is still an open question, as is closing the gap with the lower bound of Proposition~\ref{prop:AsymptoticLB}.

It would also be interesting to determine the generalized covering-radius hierarchy of known codes. These may be extreme in some cases. As we saw in Example~\ref{exemp.Hamming}, the Hamming code satisfies $R_t=t$, and in particular the covering-radius hierarchy is strictly increasing, that is,  $R_t < R_{t+1}$ for all $t \in [n-k-1]$.  This property is exclusive to the Hamming code (except other trivial cases).

\begin{proposition}
Let  $C$ be an  $[n,k\geq 1,d\geq 3]$ linear code over $\F_q$. Then 
\[R_1 < R_2 < \ldots < R_{n-k}=n-k,\]
if and only if $C$ is the  $q$-ary Hamming code.
\end{proposition}

\begin{IEEEproof}
Suppose that the covering-radius hierarchy of a code $C$ is strictly increasing. Since $R_{n-k}=n-k$,  we must have $R_1 = 1$. But a linear code with parameters $[n,k,d \geq 3]$ with covering radius $R_1 = 1$ is $1$-perfect and it must be the $q$-ary Hamming code with parameters $n = \frac{q^m-1}{q-1}$, $k = n-m$, and $d=3$.
\end{IEEEproof}

In contrast with the Hamming code, whose generalized covering radii are all distinct, the opposite occurs with MDS codes. As was shown in~\cite{gabidulin1998newton,BarGiuPla15}, the (first) covering radius of $[n,k]$ MDS codes is $n-k$, except in rare cases where it is $n-k-1$. Since the upper limit on the generalized covering radius is $n-k$, the entire hierarchy is either constant, or is a step function.

Finally, we have an algorithmic question: Given a parity-check matrix $H$ for an $[n,k]$ code over $\F_q$, and given vectors $\os_1,\dots,\os_t \in\F_q^{n-k}$, how do we efficiently find $R_t$ columns of $H$ that span the $t$ vectors? These questions, and many others, are left for future research.

\bibliographystyle{IEEEtranS}

\begin{thebibliography}{10}
\providecommand{\url}[1]{#1}
\csname url@samestyle\endcsname
\providecommand{\newblock}{\relax}
\providecommand{\bibinfo}[2]{#2}
\providecommand{\BIBentrySTDinterwordspacing}{\spaceskip=0pt\relax}
\providecommand{\BIBentryALTinterwordstretchfactor}{4}
\providecommand{\BIBentryALTinterwordspacing}{\spaceskip=\fontdimen2\font plus
\BIBentryALTinterwordstretchfactor\fontdimen3\font minus
  \fontdimen4\font\relax}
\providecommand{\BIBforeignlanguage}[2]{{%
\expandafter\ifx\csname l@#1\endcsname\relax
\typeout{** WARNING: IEEEtranS.bst: No hyphenation pattern has been}%
\typeout{** loaded for the language `#1'. Using the pattern for}%
\typeout{** the default language instead.}%
\else
\language=\csname l@#1\endcsname
\fi
#2}}
\providecommand{\BIBdecl}{\relax}
\BIBdecl

\bibitem{ashikhmin1999new}
A.~Ashikhmin, A.~Barg, and S.~Litsyn, ``New upper bounds on generalized
  weights,'' \emph{IEEE Trans.~Inform.~Theory}, vol.~45, no.~4, pp. 1258--1263,
  1999.

\bibitem{BarGiuPla15}
D.~Bartoli, M.~Giulietti, and I.~Platoni, ``On the covering radius of {MDS}
  codes,'' \emph{IEEE Trans.~Inform.~Theory}, vol.~61, no.~2, pp. 801--811,
  Feb. 2015.

\bibitem{blinovskii1990covering}
V.~M. Blinovskii, ``Covering the {H}amming space with sets translated by
  vectors of a linear code,'' \emph{Problemy Peredachi Informatsii}, vol.~26,
  no.~3, pp. 21--26, 1990.

\bibitem{chazelle1989computing}
B.~Chazelle and B.~Rosenberg, ``Computing partial sums in multidimensional
  arrays,'' in \emph{Proceedings of the fifth annual symposium on Computational
  geometry}, 1989, pp. 131--139.

\bibitem{chor1995private}
B.~Chor, O.~Goldreich, E.~Kushilevitz, and M.~Sudan, ``Private information
  retrieval,'' in \emph{Proceedings of IEEE 36th Annual Foundations of Computer
  Science}.\hskip 1em plus 0.5em minus 0.4em\relax IEEE, 1995, pp. 41--50.

\bibitem{Cohen}
G.~Cohen, I.~Honkala, S.~Litsyn, and A.~Lobstein, \emph{Covering codes}.\hskip
  1em plus 0.5em minus 0.4em\relax North-Holland, 1997.

\bibitem{cohen1985good}
G.~Cohen and P.~Frankl, ``Good coverings of {H}amming spaces with spheres,''
  \emph{Discrete Mathematics}, vol.~56, no. 2-3, pp. 125--131, 1985.

\bibitem{cohen1994upper}
G.~Cohen, S.~Litsyn, and G.~Z{\'e}mor, ``Upper bounds on generalized
  distances,'' \emph{IEEE Trans.~Inform.~Theory}, vol.~40, no.~6, pp.
  2090--2092, 1994.

\bibitem{Didier2006}
F.~Didier, ``A new upper bound on the block error probability after decoding
  over the erasure channel,'' \emph{IEEE Trans.~Inform.~Theory}, vol.~52,
  no.~10, pp. 4496--4503, Oct 2006.

\bibitem{dougherty2010higher}
S.~T. Dougherty and S.~Han, ``Higher weights and generalized {MDS} codes,''
  \emph{J. Korean Math. Soc}, vol.~47, no.~6, pp. 1167--1182, 2010.

\bibitem{1992-Feng}
G.~L. Feng, K.~K. Tzeng, and V.~K. Wei, ``On the generalized {H}amming weights
  of several classes of cyclic codes,'' \emph{IEEE Trans.~Inform.~Theory},
  vol.~38, no.~3, pp. 1125--1130, May 1992.

\bibitem{Forney1994}
G.~D. Forney, ``Density/length profiles and trellis complexity of linear block
  codes and lattices,'' in \emph{Proceedings of the 1994 IEEE International
  Symposium on Information Theory (ISIT1994), Trondheim, Norway}, June 1994, p.
  339.

\bibitem{gabidulin1998newton}
E.~M. Gabidulin and T.~Kl\o{}ve, ``The {N}ewton radius of {MDS} codes,'' in
  \emph{1998 Information Theory Workshop (ITW) Killarney, Ireland}, Jun. 1998,
  pp. 50--51.

\bibitem{Gabidulin}
E.~Gabidulin, ``Combinatorial metrics in coding theory,'' in \emph{2nd
  International Symposium on Information Theory, Armenia, USSR}, 1971.

\bibitem{GopHuaSimYek12}
P.~Gopalan, C.~Huang, H.~Simitci, and S.~Yekhanin, ``On the locality of
  codeword symbols,'' \emph{IEEE Trans.~Inform.~Theory}, vol.~58, no.~11, pp.
  6925--6934, Nov. 2012.

\bibitem{1998-Heijnen-Pellikaan}
P.~Heijnen and R.~Pellikaan, ``Generalized {H}amming weights of q-ary
  {R}eed-{M}uller codes,'' \emph{IEEE Trans.~Inform.~Theory}, vol.~44, no.~1,
  pp. 181--196, Jan 1998.

\bibitem{1992-Helleseth}
T.~Helleseth, T.~Kl\o{}ve, and {\O{}}.~Ytrehus, ``Generalized {H}amming weights
  of linear codes,'' \emph{IEEE Trans.~Inform.~Theory}, vol.~38, no.~3, pp.
  1133--1140, May 1992.

\bibitem{HoBruAgr98}
C.-T. Ho, J.~Bruck, and R.~Agrawal, ``Partial-sum queries in {OLAP} data cubes
  using covering codes,'' \emph{IEEE Trans.~Comput.}, vol.~47, no.~12, pp.
  1326--1340, Dec. 1998.

\bibitem{Feng}
L.~X. K.~Feng and F.~J. Hickernell, ``Linear error-block codes,'' \emph{Finite
  Fields and Their Applications}, vol.~12, pp. 638--652, 2006.

\bibitem{KLOVE1992}
T.~Kl{\o}ve, ``Support weight distribution of linear codes,'' \emph{Discrete
  Mathematics}, vol. 106-107, pp. 311 -- 316, 1992.

\bibitem{Kurihara}
J.~Kurihara, R.~Matsumoto, and T.~Uyematsu, ``Relative generalized rank weight
  of linear codes and its applications to network coding,'' \emph{IEEE
  Trans.~Inform.~Theory}, vol.~61, no.~7, pp. 3912--3936, July 2015.

\bibitem{Lemes}
L.~C. Lemes and M.~Firer, ``Generalized weights and bounds for error
  probability over erasure channels,'' in \emph{2014 Information Theory and
  Applications Workshop (ITA), San Diego, CA, USA}, Feb 2014, pp. 1--8.

\bibitem{oggier}
F.~Oggier and A.~Sboui, ``On the existence of generalized rank weights,'' in
  \emph{2012 International Symposium on Information Theory and its Applications
  (ISITA), Honolulu, HI, USA}, Oct 2012, pp. 406--410.

\bibitem{potukuchi2020improved}
A.~Potukuchi and Y.~Zhang, ``Improved efficiency for covering codes matching
  the sphere-covering bound,'' in \emph{2020 IEEE International Symposium on
  Information Theory (ISIT)}.\hskip 1em plus 0.5em minus 0.4em\relax IEEE,
  2020, pp. 102--107.

\bibitem{TamWanBru14}
I.~Tamo, Z.~Wang, and J.~Bruck, ``Access versus bandwidth in codes for
  storage,'' \emph{IEEE Trans.~Inform.~Theory}, vol.~60, no.~4, pp. 2028--2037,
  Apr. 2014.

\bibitem{tsfasman1995geometric}
M.~A. Tsfasman and S.~G. Vladut, ``Geometric approach to higher weights,''
  \emph{IEEE Trans.~Inform.~Theory}, vol.~41, no.~6, pp. 1564--1588, 1995.

\bibitem{1991-Wei}
V.~K. Wei, ``Generalized {H}amming weights for linear codes,'' \emph{IEEE
  Trans.~Inform.~Theory}, vol.~37, no.~5, pp. 1412--1418, Sep. 1991.

\bibitem{ZhaYaaEtzSch19}
Y.~Zhang, E.~Yaakobi, T.~Etzion, and M.~Schwartz, ``On the access complexity of
  {PIR} schemes,'' in \emph{Proceedings of the 2019 IEEE International
  Symposium on Information Theory (ISIT2019), Paris, France}, Jul. 2019, pp.
  2134--2138.

\end{thebibliography}

\end{document}